\documentclass{article}
\usepackage[utf8]{inputenc}

\usepackage{physics}

\usepackage{graphicx}
\usepackage{hyperref}
\usepackage{amsmath}
\usepackage{amssymb}
\usepackage{tikz}
\usepackage{blindtext}
\usepackage{geometry}
\usepackage{caption}
\usepackage{subcaption}

\usepackage{comment}

\usepackage{float}

\usepackage[bottom]{footmisc}

\usepackage{abstract}

\usetikzlibrary{quantikz}

\newcommand{\tcr}[1]{\textcolor{red}{#1}}

\usepackage{mathdots}

\usepackage{rotating}

\usepackage{enumitem}

\usepackage[affil-it]{authblk}

\graphicspath{{./Figures/}}

\makeatletter
\def\@maketitle{
  \begin{center}%
  \let \footnote \thanks
    {\LARGE\bfseries \@title \par}%
    \vskip 1.5em%
    {\large
      \lineskip .5em%
      \begin{tabular}[t]{c}%
        \@author
      \end{tabular}\par}%
  \end{center}%
  \par
  \vskip 1.0em}
\makeatother

\begin{document}

\title{An introduction to financial option pricing on a qudit-based quantum computer}

\author{Nicholas Bornman%
    \thanks{Email: nbornman@fnal.gov}}
\affil{{\small Superconducting Quantum Materials and Systems Center (SQMS), Fermi National Accelerator Laboratory, Batavia, IL 60510, USA}}


\maketitle

\begin{abstract}

The financial sector is anticipated to be one of the first industries to benefit from the increased computational power of quantum computers, in areas such as portfolio optimisation and risk management to financial derivative pricing. Financial mathematics, and derivative pricing in particular, are not areas quantum physicists are traditionally trained in despite the fact that they often have the raw technical skills needed to understand such topics. On the other hand, most quantum algorithms have largely focused on qubits, which are comprised of two discrete states, as the information carriers. However, discrete higher-dimensional qudits, in addition to possibly possessing increased noise robustness and allowing for novel error correction protocols in certain hardware implementations, also have logarithmically greater information storage and processing capacity. In the current NISQ era of quantum computing, a wide array of hardware paradigms are still being studied and any potential advantage a platform offers is worth exploring. Here we introduce the basic concepts behind financial derivatives for the unfamiliar enthusiast as well as outline in great detail the quantum algorithm routines needed to price a European option, the simplest derivative. This is done within the context of a quantum computer comprised of qudits and employing the natural higher-dimensional analogue of a qubit-based pricing algorithm with its various subroutines. From these pieces, one should relatively easily be able to tailor the scheme to more complex, realistic financial derivatives. Finally, the entire stack is numerically simulated with the results demonstrating how the qudit-based scheme's payoff quickly approaches that of both a similarly-resourced classical computer as well as the true payoff, within error, for a modest increase in qudit dimension.

\end{abstract}

\section{Financial derivatives: a primer's primer}
\label{sec:classfin}

Financial assets can take a variety of forms, such as company stocks, government bonds, owning the debt of mortgage holders, as well as raw commodities such as precious metals and important foodstuffs. In reality, many of these assets are not themselves exchanged between a buyer and seller on the market. Rather, financial derivatives, which are simply contracts that are based on one or more of these underlying assets, are entered into and are the items that are in fact traded. The contracts themselves have values `derived' from the underlying assets, and given that the assets' values are subject to various - largely indeterministic - market forces and other processes, the prime focus of derivative pricing in financial mathematics is to determine the fair value of such contracts.

The simplest kind of derivative is called a \textit{futures} contract. A future is a publicly-traded contract which obliges the buyer to purchase a particular asset from the seller at a pre-agreed upon price at a later date. The utility of such a contract is that it allows investors either to make money by speculating (which is a risky investment strategy focused on taking advantage of price fluctuations to earn a profit), or to hedge (i.e. reduce the risk of) their betting position against market fluctuations. For example, if one speculates that the price of an asset is likely to increase at a later date, one could, today, purchase a futures contact in which one agrees to purchase the asset at the later date at a price lower than the speculated future price. On the other hand, to hedge their market position, an investor may purchase a futures contract since it allows them to lock in the purchase price of the asset, thereby creating a degree of certainty and controlling risk. However, buying or selling such contracts is never without risk: if a speculative investor purchases a future in the belief that an asset's price will increase, but the asset's price in fact decreases at the contract's maturation date, the investor will lose money as he will be forced to purchase the cheaper asset for more than its actual value. As such, the decision of whether to buy or sell any derivative should take into account the expected return of investing in the most `risk-free' investment - typically US government treasury bonds - available on the market.

Financial \textit{options} contracts, on the other hand, are similar to futures, but with one key difference: they give the buyer the prerogative, not the obligation, to either buy or sell the asset. In particular, the simplest option, a European call (put) option, gives the buyer the right to purchase (sell) the asset (which we take to have a value of $S_t$ at time $t$, with today being $t=0$), at the strike price $K$ once the contract is exercised at time $t=T$. If the asset's price at the expiration date $t=T$ is $S_T$, the payoff (denoted by $f$) for the buyer of the call option would be

\begin{equation}
f(S_T) = \max(0, S_T - K).
\label{eqn:payoff}
\end{equation}

This formula is intuitive: if it ends up being that $S_T > K$ at expiration, it makes sense for the buyer to exercise the contract as he'll be acquiring an asset with a value of $S_T$ for the cheaper price of $K$, and hence making a profit of $S_T - K$. However, if $S_T < K$, the buyer should not exercise the right to buy the asset since its mandated purchase price of $K$ would be more than its actual market value.

\begin{figure}[h]
    \centering
    \includegraphics[width=0.65\textwidth]{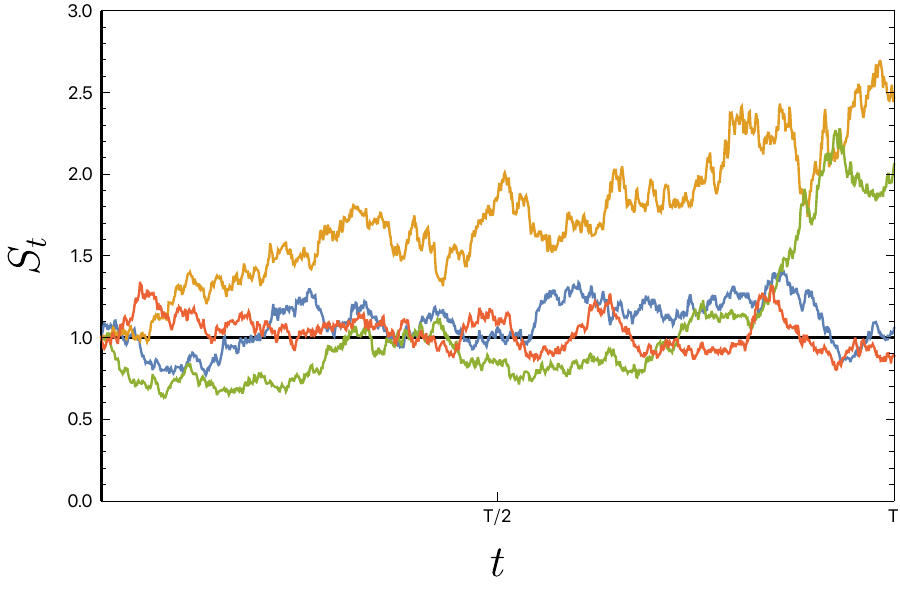}
    \caption{A sample of stochastic asset price paths (modelled using geometric Brownian motion with drift $0.05$ and volatility $0.2$, c.f. Eqn. \ref{eqn:BSMmodel}). The market's inherent indeterminism means that we don't know which path the asset's price will follow over time; we can only compute the expected payoff, which is `averaged' over all possible paths under the chosen stochastic model.}
    \label{fig:stochastic}
\end{figure}

If one somehow knew, today, that the asset's price would be $S_T$ at time $T$, the fair price of a European call option based on that asset would be the discounted future payoff, namely $e^{-rT}f(S_T)$. Here, $r$ is the continuously-compounded return rate of risk-free government bonds, and the $e^{-rT}$ discount factor accounts for the opportunity cost of not having invested in these bonds instead of the asset. However, asset prices do not evolve deterministically, but stochastically (see Fig. \ref{fig:stochastic}). Indeed, the value of a single asset, such as the price of a barrel of oil or a share of a publicly-listed company, is influenced by a number of factors such as reserve bank interest rates, traditional supply and demand economics, as well as global geopolitics and human psychology \cite{ammendola2000devil}. Furthermore, the global market is comprised of countless assets, all influencing one another in a terribly interconnected, unpredictable fashion. This indeterminism complicates matters so much that one cannot compute the exact future value of an asset, let alone an exotic option based on it. Instead, we use the discounted, \textit{expected} payoff, namely $e^{-rT}\text{E}_{\text{P}}[f(S_T)]$. This expectation value is taken with respect to a particular probability distribution $\text{P}$ of $S_t$ (we'll elaborate on the distribution shortly). 

During early attempts to understand the complicated problem of option pricing, one of the first - and most important - option pricing models proposed was the Black-Scholes-Merton (BSM) model \cite{bsmodel}. It takes the asset's price to be primarily steered by the most common stochastic processes: one-dimensional Brownian motion (see \cite{baz2004financial} for an intuitive primer on Brownian motion). If $W_t$ denotes such a Brownian process and $S_t$ the price of the asset at time $t \leq T$, the BSM model assumes that $S_t$ follows this stochastic differential equation (known as geometric Brownian motion)

\begin{equation}
dS_t = \alpha S_t dt + \sigma S_t dW_t.
\label{eqn:BSMmodel}
\end{equation}

Here, the asset's evolution comprises two parts: the first term models a deterministic trend, or drift, in the `mean' price of the asset (with $\alpha$ the percentage drift), whereas the second term (with $\sigma$ the percentage volatility and $dW_t$ a Brownian increment) models the unpredictable changes in the asset's price during each time increment. In the BSM model, the drift and volatility are assumed to be constants, independent of the asset price $S_t$ and time (although more realistic stochastic volatility models, in which $\sigma$ itself is driven by a separate Brownian process, exist \cite{hull2003options}). Finally, it should be noted that in a formal treatment of stochastic calculus, a process such as $W_t$ would need to be considered with respect to a probability measure. There are a number of such measures with respect to which $W_t$ models standard Brownian motion. However, such a formal treatment is beyond the scope of this introduction.

$W_t$ is stochastic and has a quadratic variance proportional to the time increment $t$, and ordinary calculus cannot solve Eq. \ref{eqn:BSMmodel} for $S_t$ as a function of $t$ and $W_t$. However, using the famed It\^{o}'s lemma (see \cite{hull2003options}), we can solve for $S_t$ as,

\begin{equation}
S_t = S_0 e^{\sigma W_t + (\alpha - \frac{\sigma^2}{2})t},
\label{eqn:Stprice}
\end{equation}

\noindent
where $S_0$ is the asset's price at $t=0$. Fig. \ref{fig:stochastic} represents samples of this function.

Eq. \ref{eqn:Stprice} gives, in an informal sense, a relation between $S_t$ and one particular \textit{realisation} of the Brownian motion $W_t$. One cannot definitively say what value $S_t$ will assume at a later stage because one cannot definitively say what value $W_t$ will assume: $W_t$ has independent, random increments. Even so, these increments have a well-defined distribution: one property characterising $W_t$ is that $W_t - W_0$ is normally distribution with mean $0$ and variance $t$. This suggests that $S_t$ itself potentially has a well-defined distribution. Indeed, re-writing Eq. \ref{eqn:BSMmodel} as $\frac{dS_t}{S_t} = \alpha dt + \sigma dW_t$, the process $\frac{dS_t}{S_t}$ is an It\^{o} drift-diffusion process comprising a fixed drift term and a random volatile term. Given this, the Fokker-Planck equation \cite{fokker1914mittlere, planck1917satz} is a partial differential equation describing the evolution of the probability density of a process subject to these two exact kinds of forces. Solving it (see \cite{Risken1996}) for the probability density $p(S_t; t, \alpha, \sigma, S_0)$ gives

\begin{equation}
p(S_t; t, \alpha, \sigma, S_0) = \frac{1}{S_t \sigma \sqrt{2 \pi t}} \exp(\frac{-\left(\ln(\frac{S_t}{S_0}) - (\alpha - \frac{1}{2}\sigma^2) t \right)^2}{2\sigma^2 t}),
\label{eqn:probdensSt}
\end{equation}

\noindent
i.e. the asset's price evolution follows a log-normal distribution.

We almost have all of the ingredients needed to compute the realistic discounted expected payoff for the European call option of Eq. \ref{eqn:payoff}. However, we need to hit all of this mathematics with a dose of real-world financial economics. Suppose that we start with the assumption that the market is rational (despite much evidence to the contrary \cite{investorirrationality} - as noted economist John Maynard Keynes opined: ``The market can stay irrational longer than you can stay solvent''). One characteristic of a rational market is the assumption that, for a given capital investment, one cannot make a profit larger than the profit from having instead invested in a risk-free asset, without taking on risk. Such a market is called \textit{arbitrage-free} and is based on the presumption that if an arbitrage opportunity were present in reality, market forces would speedily move to eliminate the opportunity\footnote{For example, if it were possible to purchase a commodity at a lower price on one exchange and immediately sell it on another at a higher price (with negligible fees), many people would likely exploit this opportunity and earn profit, with no risk. The lower-priced market would likely quickly become aware of the disparity and consequently increasing the commodity's price (so as not to make a loss itself), thereby eliminating the arbitrage opportunity.}. In this way, the market moves towards an arbitrage-free equilibrium.

With this in mind, the first fundamental theorem of asset pricing states that a market is arbitrage-free if and only if there exists a so-called \textit{risk-neutral} measure \cite{hull2003options}. Intuitively, such a measure P is one with respect to which the spot price (i.e. today's price) of an asset is exactly equal to the expected future price of the asset discounted to today, namely $S_0 \equiv e^{-rt}\text{E}_{\text{P}}[S_t]$, for all $t$. While we will not delve into such concepts further, a formal, in-depth definition of derivative pricing and measures is best understood in terms of measure-theoretic probability spaces \cite{roussas2014introduction}. However, in the result, choosing a risk-free neutral measure in the current context is equivalent to setting the drift $\alpha$ of the underling asset to identically equal $r$, the risk-free interest rate\footnote{Indeed, if this weren't the case, an arbitrage opportunity would exist: if $\alpha < r$, one could construct a portfolio consisting of short positions of the asset $S_t$; if $\alpha > r$, one could purchase the asset $S_t$, hold it, and sell it later for more than the purchase price. In this way, one would be guaranteed to, `on average', make a profit \cite{PhysRevA.98.022321}.}. With $p(S_t; t, \alpha = r, \sigma, S_0)$ from Eq. \ref{eqn:probdensSt} and the payoff function $f$ from Eq. \ref{eqn:payoff}, we can now compute the European call option's fair value, denoted $\nu$, according to

\begin{equation}
\nu = e^{-rT}\text{E}_{\text{P}}[f(S_T)].
\label{eqn:optionpricingproblem}
\end{equation}

An analytic expression for the above quantity exists (see \cite{PhysRevA.98.022321}), given the relative simplicity of pricing a single asset whose payoff depends on only a single time point, namely the contract expiration date $T$. Unsurprisingly, a whole host of options far more complex than European options exist. Examples include American options (whereby one can exercise the option at \textit{any} time prior to the expiration date of the contract) and Asian options (in which the payoff $f$ doesn't depend merely on the asset's price at the contact's expiration date, but rather upon the asset's average price $\overline{S}$ sampled at regular intervals over the contract length). Worth mentioning, too, are barrier options. In the case of `knock-out' barrier options, the payoff $f$ at the maturation date $t = T$ is $0$ if the asset's price $S_t$ (with $t < T$) ever crosses a pre-defined `barrier' value $B$ at any time before the contract matures (and if it doesn't, the payoff follows Eq. \ref{eqn:payoff}); for `knock-in' barrier options, the payoff is $0$ unless the asset's price crosses the barrier at least once. Finally, one can string many of these options and concepts together in increasing complexity. For example, options contracts with a payoff function with multiple barriers are commonplace \cite{chakrabarti2021threshold}, as are options consisting of baskets of assets rather than only one. While we only discuss the mathematical details of European options in this text, it is clear that things can quickly get out of hand when trying to price real-world financial derivatives.

\section{Qudit option pricing algorithm}
\label{sec:quditalg}

Many real-world options depend not only on the asset's final price, $S_T$, but rather on the asset's value $S_t$ at a number of time points, $0 < t \leq T$. In other words, the \textit{path} the asset's price follows, not merely the path's endpoint. Such path-dependent option pricing calculations are expensive and investment firms typically carry out such computations overnight, on high-performance classical computers using Monte Carlo integration techniques\footnote{\label{fnt:classmontecarlo}To classically price complex options for which nice analytic solutions do not exist, a large number $M$ of potential asset price paths, $\{ \mathbf{S}_i \}_{i=1}^M$ (where $\mathbf{S}_i$ is a single realisation of the asset's value evolution over the length of the contract), are first sampled from the underlying probability distribution $\text{P}$ (which can be either known, or implied by market data). An estimator $\hat{\text{E}}_{\text{P}}[f]$ of the true expected payoff $\text{E}_{\text{P}}[f]$ is then constructed by simply taking the average over the paths, i.e.

\begin{equation}
\hat{\text{E}}_{\text{P}}[f] = \frac{1}{M} \sum_{i = 1}^M f(\mathbf{X}_i).
\label{eqn:classicalmontecarlo}
\end{equation}

Finally, from the central limit theorem, as $M$ grows, the difference between $\hat{\text{E}}_{\text{P}}[f]$ and $\text{E}_{\text{P}}[f]$ converges as $\order{M^{-1/2}}$ \cite{caflisch1998monte}}. It is \textit{path-dependent} derivative pricing that is expected to gain the most from quantum computers \cite{chakrabarti2021threshold}. However, our goal here is not to discuss the pricing of such complex options on a qudit-based quantum computer. Our goal is first to give an introduction to some of the concepts behind financial option pricing and work through the mathematical details, from start to finish, of how a simple option would be priced on a qubit-based quantum computer. As such, we will primarily restrict ourselves to a European call option and outline the protocol in great detail. Simultaneously, our second goal is to simulate the analogous protocol on a quantum computer with an information register comprised of qudits. While we don't expect to see any advantage beyond the expected logarithmic increase in information storage for a qudit register in the completely analogous qudit protocol, qudit-based quantum computers offer a number of advantages over their qubit counterparts \cite{wang2020qudits, chi2022programmable} and any future implementation of derivative pricing on such a machine would need to begin with just such an analysis as this.

The traditional quantum derivative pricing algorithm is based on quantum Monte Carlo integration (see, for example, \cite{montanaro2015quantum}), which is an extension of the quantum amplitude estimation (QAE) algorithm \cite{brassard2002quantum} to a weighted combination of amplitudes. The scheme is relatively simple. First, one prepares a `probability' register of qubits into a weighted superposition of states. Each basis state in the superposition represents a potential asset price $S_T$, and the weighted coefficient represents the corresponding probability of realising $S_T$. Second, a series of controlled entangling gates is applied, with the probability register functioning as the control, to load the payoff function into the amplitude of a target ancilla qubit's $\ket{1}$ basis state. As such, this amplitude becomes a weighted combination of payoffs, with the weights being the probabilities of the payoffs themselves. This is akin to classically integrating the payoff over the probability density. And finally, QAE is used to estimate the amplitude of the ancilla's $\ket{1}$ basis state. The resulting quantity can easily be classically mapped to the expected payoff $\text{E}_{\text{P}}[f(S_T)]$ and the European option's fair value computed (as discounting $\text{E}_{\text{P}}[f(S_T)]$ is computationally cheap).

The chief advantage of pricing options using this algorithm on quantum computers rather than employing high performance classical resources is due to the quadratic speedup inherent in the quantum algorithm. Indeed, as outlined in footnote \ref{fnt:classmontecarlo}, the discrepancy between the true expected payoff and the estimator that classical Monte Carlo integration outputs, scales as $\order{M^{-1/2}}$. The more samples paths $M$ taken, the smaller the error. However, this convergence is not particularly fast: if one takes 100 times as many samples (i.e. $M \to 100M$), one gains only an order of magnitude of accuracy ($M^{-1/2} \to 10M^{-1/2}$). In the quantum algorithm to be outlined, however, this discrepancy scales as $\order{M^{-1}}$ \cite{brassard2002quantum}, where $M$ here is the quantum analogue of the number of samples taken (in particular, $M$ in the quantum case is the number of calls to a particular operator, called an oracle, which we shall discuss later). This is the much-touted speedup quantum computers provide.

Before the qudit extension of this algorithm is outlined in detail and simulated below, it is worth mentioning that the `phase estimation' subroutine in QAE currently presents the biggest bottleneck in experimental implementations of quantum derivative pricing \cite{kitaev1995quantum, nielsen2010quantum}. However, it has been numerically demonstrated that one can still asymptotically achieve the quadratic speedup by forgoing the phase estimation subroutines and instead employing maximum-likelihood estimation \cite{suzuki2020amplitude}. This is also the case in a qudit-based algorithm, and as such, we shall discuss and implement this version of amplitude estimation without phase estimation.

\subsection{Step 1: Load the distribution}
\label{subsec:probloading}

Suppose that we have access to a quantum system comprised of $n$ fully-connected qudits, each of which lies in a $d$-dimensional Hilbert space (such systems may arise from, for example, the linear modes of superconducting cavities coupled to transmon qubits \cite{alam2022quantum}). The computational basis states of qudit $j$, where $j \in \{0, 1, \cdots, n-1 \}$, are $\{ \ket{0}_j, \ket{1}_j, \cdots, \ket{d-1}_j \}$. As such, the composite Hilbert space $\mathcal{H}$ is spanned by the computational basis states of the form $\ket{i}_{(n)} = \ket{i_0}_0 \otimes \ket{i_1}_1 \otimes \cdots \otimes \ket{i_{n-1}}_{n-1}$. Note that $i \in \{0, 1, \cdots, d^n - 1\}$, expressed in base-$d$, is $i = i_0 + d^1i_1 + \cdots + d^{n-1}i_{n-1}$, and hence $i$ holds $\log_2 d^n$ bits of information.

Our first task is to create an operator $\hat{P}$ which loads the probability distribution of the asset $S_T$ into the register of qudits. This probability distribution could be derived from either a financial model or implied from market data the modeller has at hand. For our single asset in the BSM model, the distribution is $p(S_T; T, r, \sigma, S_0)$ in Eq. \ref{eqn:probdensSt} (we will suppress the parameters and denote this distribution by $p(S_T)$ from here onwards).

The asset price $S_T$ can theoretically assume any non-negative real value. However, a finite register cannot faithfully represent this uncountable set (if the probability distribution for the asset were discrete, the situation would simplify slightly). So, we first truncate the asset price domain $[ 0, \infty)$ to $\text{S} = [S_{\text{min}}, S_{\text{max}}]$ by choosing $S_{\text{min}}$ and $S_{\text{max}}$ such that most of the probability mass lies within the truncated region\footnote{This assumption necessarily introduces an error into the calculations as unlikely but `extreme' events, which reside within the tails of the distribution, are discarded. The error introduced should be acceptable for short-tailed distributions, but it should be noted that potential extreme, unlikely events still affect the market in practice \cite{taleb2007black}.}. The choice of this truncation is ultimately a design choice influenced by the desired accuracy, the distribution itself and available register size. However, for an underlying distribution $\text{P}$ with mean $\mu$ and standard deviation $\sigma$, it is reasonable to set $S_{\text{min}} = \max(0, \mu - 3\sigma)$ and $S_{\text{max}} = \mu + 3\sigma$ to capture $3 \sigma$ worth of data.

Next, we approximate $\text{P}$ by sampling it at discrete points in $\text{S}$, since our qudit register can only hold $d^n$ values. This is done by first dividing the region $\text{S}$ into $d^n$ partitions of equal length $\omega$, where $\omega = (S_{\text{max}} - S_{\text{min}})/d^n$. Sampling $\text{S}$ at the centre of each of these intervals gives a set $\{ s_i \}$, where

\begin{equation}
s_i = S_{\text{min}} + \left(i + \frac{1}{2} \right)\omega, \; \text{with} \; i \in \{0, \cdots, d^n-1\}.
\label{eqn:affinemap}
\end{equation}

Eq. \ref{eqn:affinemap} provides the affine transformation between the integers $i$ stored in the qudit register and the asset values $\{ s_i \}$ they represent. This mapping will become important later.

Next, the distribution $\text{P}$ is discretised by sampling it at points $\{ s_i\}$. Finally, a set of probability values $\{ p_i \}$, which will represent $\text{P}$, is formed by normalising the samples of $\text{P}$, i.e. $p_i = \frac{1}{\mathcal{N}} \times p(s_i)$ with $\mathcal{N} = \sum_{i=0}^{d^n-1} p(s_i)$. In a rough sense, $p_i$ is the chance of the asset's value assuming a value `close to' $s_i$ at the expiration date $T$.

With this setup, the first step of the qudit option pricing scheme is:

\begin{itemize}
\item[1] Construct a unitary operator $\hat{P}$ which loads the probabilities $\{ p_i \}$ into the coefficients of a register of qudits initialised in the ground state, i.e.

\begin{equation}
\ket{0}_{(n)} \xrightarrow{\hat{P}} \sum_{i} \sqrt{p_i} \ket{i}_{n} = \hat{P}\ket{0}_{(n)}.
\label{eqn:Poperator}
\end{equation}
\end{itemize}

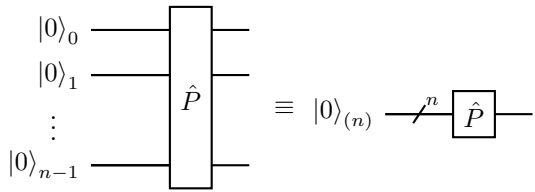
\begin{figure}[h]
    \centering
    \begin{quantikz}[align equals at=2.5, row sep={0.6cm,between origins}]
        \lstick{$\ket{0}_{0}$} & \gate[wires=4,nwires={3}]{\hat{P}} & \qw \\
        \lstick{$\ket{0}_1$} & \qw & \qw \\
        \vdots \hspace{1cm} && \\
        \lstick{$\ket{0}_{n-1}$} & \qw & \qw
    \end{quantikz}
    \; $\equiv$ \begin{quantikz}
    \lstick{$\ket{0}_{(n)}$} &[4mm] \gate[]{\hat{P}} \qwbundle{n} & \qw
    \end{quantikz}
    \caption{Probability loading subroutine}
    \label{circ:probload}
\end{figure}

Creating a general state preparation subroutine $\hat{P}$ (Fig. \ref{circ:probload}) often requires very large numbers of fundamental gates (see Sec. \ref{subsubsec:resest}). Whatever our choice for the unitary $\hat{P}$ however, the state preparation step above only requires that the first column in the matrix representation of $\hat{P}$, denoted $\mathbb{P}$, be equal to $\mathbf{p} = [\sqrt{p_0}, \sqrt{p_1}, \cdots, \sqrt{p_{d^n-1}}]^{\text{T}}$ (up to permutation, to match the chosen basis). While many potential choices of $\mathbb{P}$ with this property exist, one simple unitary matrix with the requisite first column can be constructed using Householder transformations \cite{Householder}: define $\mathbf{w} = \mathbf{p} - \mathbf{e}_0$ (with $\mathbf{e}_0 = [1, 0, 0, \cdots, 0]^{\text{T}}$), and note that $\mathbf{w}^{\dagger}\mathbf{w} = 2(1-\sqrt{p_0})$. Then, a $\mathbb{P}$ with the correct first column is given by

\begin{equation}
\mathbb{P} = \mathbb{I} - \frac{\mathbf{w} \; \mathbf{w}^{\dagger}}{1-\sqrt{p_0}}.
\label{eqn:Pmatrixrep}
\end{equation}

It can easily be verified that $\mathbb{P} \, \mathbf{e}_0 = \mathbf{p}$, which is the matrix representation equivalent of Eq. \ref{eqn:Poperator}. We use this choice of $\hat{P}$ in the forthcoming simulations.

\subsubsection{Resource comments for \texorpdfstring{$\hat{P}$}{P} (and general unitary gates)}
\label{subsubsec:resest}

Given the relative dearth of studies of quantum computers based on qudits, both theoretical and experimental, and what their fundamental gate set(s) would be (with which one must decompose the state one wishes to prepare, into), it is difficult to give accurate resource estimates for arbitrary state preparation as well as unitary compilation. Furthermore, one cannot make a meaningful `apples to apples' gate resource comparison between qubit and qudit architectures without clearly analogous fundamental gate sets. However, some important comments can nonetheless be made.

Arbitrary state preparation and unitary compilation, using a conventional straightforward `decomposition' scheme, on an $n$ qubit gate-based system with a fundamental gate set comprised of arbitrary single-qubit gates and CNOT gates, generally requires $O(a^n)$ single qubit gates as well as $O(a^n)$ CNOT gates (where $a = 2$ in the case of state preparation \cite{plesch2011quantum, mottonen2004transformation} and $a = 4$ for unitary compilation \cite{vartiainen2004efficient, mottonen2004quantum}). This exponential scaling renders probability encoding schemes, via state preparation as above, infeasible for large $n$. This is because the current NISQ era of qubit-based quantum computing (which is characterised by imperfect gates, short qubit coherence times and no algorithmic error correction \cite{preskill2018quantum}) precludes algorithms with many operations and/or long execution times. These same concerns arise in a qudit-based computer comprised of the natural qudit analogues of the well-known $X$, $Z$ and $CZ$ gates (see \cite{luo2014universal, wang2020qudits} for details). Indeed, decomposing an arbitrary unitary gate $\text{U} \in \text{SU}(d^n)$, given $n$ $d$-dimensional qudits and equipping the system with the aforementioned analogue qudit gates, requires $\mathcal{O}(nd^n)$ primitive operations \cite{luo2014universal}, which is still a prohibitive scaling. 

For the particular case, however, where a state encodes a \textit{classically efficiently-integrable} distribution (such as the log-concave BSM model, Eq. \ref{eqn:probdensSt}) in a register of qubits (and qudits), state preparation can be efficiently accomplished \cite{grover2002creating}. But, in any event, the efficient loading of log-normal distributions requires pre-computing integrals of the probability distribution - often using classical Monte Carlo integration - whereas the crux of employing quantum amplitude estimation-based option pricing in the first place was to avoid the need for classical integration. Furthermore, the underlying distributions of more realistic financial models which capture important market dynamics which the BSM model does not, or distributions suggested purely by market data, may not have analytic representations at all. In these cases, the distributions may need to be encoded using other, contemporary techniques. With this in mind, it almost goes without saying that machine learning and neural networks are becoming increasingly important tools in classical finance to capture highly complex and inter-connected market dynamics efficiently (see, for example, \cite{dixon2020machine, liu2019pricing}), as well as model the market without assuming an underlying financial mathematical model \cite{horvath2021deep}. The same holds true in the quantum context. Indeed, in \cite{zoufal2019quantum}, the authors outlined how a quantum version of a Generative Adversarial Network \cite{goodfellow2014generative} can be used to create a circuit encoding the underlying distribution of training data samples by tuning the parameters of a parameterised model ansatz circuit. The parameters are tuned according to the feedback from a classical discriminator which is fed both the training data as well as the corresponding data generated from measurements of the parameterised quantum circuit itself. The upshot of such a protocol is the freedom to choose a polynomial-length parameterised ansatz circuit which can be efficiently deployed on the available quantum computing platform \cite{Stamatopoulos2020optionpricingusing}. While incorporating neural network and machine learning techniques into quantum computing is still at the relative cutting edge of qubit-based systems, such techniques hold enormous promise and are almost certain to eventually feature in qudit-based systems too.

Finally, it is foolish to completely separate theoretical discussions of quantum circuit compilation from hardware considerations. The $X$, $Z$ and $CZ$ qudit gate analogues mentioned above naturally lend themselves to linear ion-trap \cite{kielpinski2002architecture, wang2020qudits} and multi-port photonics \cite{knill2001scheme, kok2007linear} quantum computing systems. As such, future commercial quantum computers based on such hardware would likely have access to these fundamental gate operations. On the other hand, circuit QED systems \cite{blais2021circuit} and/or microwave cavities \cite{haroche2020from} are a novel technology which also hold much promise, as they have the potential for very long coherence times \cite{romanenko2020three}, relatively efficient universal control \cite{fosel2020efficient} and error correction protocols \cite{rajeev2023suppressing}. A popular potential gate set for cavity systems is one comprised of so-called SNAP gates along with cavity displacements \cite{krastanov2015universal, fosel2020efficient}, which are intuitively different operations from qubit gates. SNAP and displacement gates are universal in that they can implement any arbitrary operation on the cavity qudit/transmon system, and importantly, if one wishes to address the first $N$ Fock states of the cavity mode, the number of SNAP and displacement gate pairs scales as $\mathcal{O}(N)$ \cite{fosel2020efficient}. This potential linear scaling of the number of gates required to affect a chosen $N$-dimensional chosen unitary could offer a potential boon for microwave cavity based systems. As always though, much more work remains to be done to fully understand such systems.

\subsection{Step 2: Compute the payoff}
\label{subsec:payoff}

Suppose that we have constructed the operator $\hat{P}$, which loads the random variable (namely the asset price $S_T$ at time $T$) into the register of qudits. We next move on to the subroutine which computes the payoff for the European call option. This payoff, Eq. \ref{eqn:payoff}, is $f(S_T) = \max(0, S_T - K)$ for a predetermined, classical strike price $K$ (with $K \in [ S_{\min}, S_{\max} ]$), and will be loaded into the amplitude of a `payoff' ancilla qubit\footnote{One could use a qudit instead of a qubit, but the most natural qudit analogue of the ordinary qubit option pricing protocol does not need the higher-dimensional nature of a potential payoff qudit.}. The subroutine for this part of the option pricing protocol comprises two parts: a part which first compares the integer representation of $S_T$ (i.e. $i$) with the integer representation of $K$ (which we call $k$), and then a part which loads either $0$ or $S_T - K$ into the ancilla's amplitude, depending on the result of this comparison\footnote{For a simpler payoff, such as that of a futures derivative (for which $f(S_T) = S_T - K$), this need for a comparison would fall away and result in a simpler payoff circuit subroutine.}. In both the qubit \cite{Stamatopoulos2020optionpricingusing} as well as the current qudit option pricing schemes, these comparison-payoff encoding steps happen `in parallel' for all asset prices $\ket{i}_{(n)}$ in the superposition state of Eq. \ref{eqn:Poperator}. Finally, note that $k$ can easily be found using the mapping in Eq. \ref{eqn:affinemap}: with $K = S_{\text{min}} + (k^{*} + \frac{1}{2})\omega$ for some $k^{*}$, $k$ is found by rounding $k^{*}$ to the nearest integer. $k$, an integer in $\{0, 1, \cdots, d^n - 1\}$, has a base-$d$ representation of $k = k_0 + d^1k_1 + d^2k_2 + \cdots + d^{n-1}k_{n-1}$.

The first `part' needed in the subroutine is an operator which compares the classical variable $k$ with $i$ from the qudit basis state $\ket{i}_{(n)}$ and then changes the state of a `comparator' ancilla qubit depending on which integer is greater. Specifically:

\begin{itemize}
\item[2.1] Create an operator $\hat{C}_k$ which flips the state of an ancilla qubit $\ket{0}_c$ depending on the computational basis state of the qudit register, according to

\begin{equation}
\ket{i}_{(n)} \ket{0}_c \xrightarrow{\hat{C}_k}
\begin{cases} 
      \ket{i}_{(n)} \ket{0}_c & \text{if} \; i < k, \\
      \ket{i}_{(n)} \ket{1}_c & \text{if} \; i \geq k. \\
\end{cases}
\label{eqn:Coperator}
\end{equation}

\end{itemize}

Once such a circuit for $\hat{C}_k$ has been constructed, it can be used to split the superposition state $\hat{P}\ket{0}_{(n)}$ from the first step, into two parts: one part corresponds to the cases where the asset price is less than the strike price $K$, and the other when it's greater than $K$. In other words,

\begin{equation}
\hat{P}\ket{0}_{(n)} \ket{0}_{c} = \sum_{i} \sqrt{p_i} \ket{i}_{n} \ket{0}_{c} \xrightarrow{\hat{C}_k} \sum_{i < k} \sqrt{p_i}\ket{i}_{(n)}\ket{0}_c + \sum_{i \geq k} \sqrt{p_i}\ket{i}_{(n)}\ket{1}_c.
\label{eqn:CPresult}
\end{equation}

We shall see that the circuit for $\hat{C}_k$ corresponds to a qudit version of the qubit quantum comparator (see, for example, \cite{cuccaro2004new}) and consists of a number of steps. First, the so-called $(d-1)$'s complement of $k$ is found\footnote{The method of complements, particularly finding nine's complement of an integer, is a well-known procedure in classical computing which is used, among other things, to find the difference between integers using a machine which can only perform addition (for more on arithmetic on classical computers, see \cite{koren2018computer}.}. Next, the integer $1$ is added to this complement, giving a string of digits we call $k^c$. Then, the sum of $k^c$ and the quantum variable $i$, which is encoded in the state $\ket{i}_{(n)}$, is computed by creating the quantum circuit equivalent of a full-adder for each individual qudit comprising the state $\ket{i}_{(n)}$ \cite{orts2020review, draper2000addition, cuccaro2004new}. A single full adder subroutine computes the sum of the $j$'th qudit state $\ket{i_j}_j$, the corresponding $j$'th digit in $k^c$, as well as any input carry digit from the previous step, and outputs the sum of these digits along with any output carry digit (which is fed into the full adder of the next step). As an example in Appendix \ref{subsec:comparator} demonstrates, adding $k^c$ to $i$ in this manner is equivalent to calculating the difference $i - k$. Finally, the overall carry digit in the classical sum of $i$ and $k^c$ is identically encoded in the state of the comparator qubit of step $2.1$ above. This qubit state hence indicates whether $k$ is larger than or equal to $i$ (in which case the comparator's state is set to $\ket{1}_c$) or not (in which case it is $\ket{0}_c$). These steps are potentially confusing, especially for those unfamiliar with classical computing arithmetic. However, the logic behind these steps as well as potential gate representations for the subroutine $\hat{C}_k$, are outlined in great detail in Appendix \ref{subsec:comparator}.

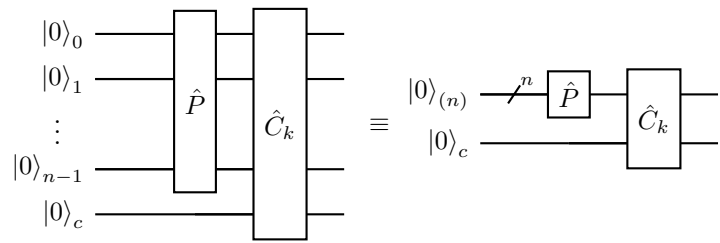
\begin{figure}[h]
    \centering
    \begin{quantikz}[align equals at=3, row sep={0.6cm,between origins}]
        & \lstick{$\ket{0}_{0}$} & \gate[wires=4,nwires={3}]{\hat{P}} & \gate[wires=5,nwires={3}]{\hat{C}_k} & \qw \\
        & \lstick{$\ket{0}_1$} & \qw & \qw & \qw \\
        & \vdots \hspace{1cm} & & & \\
        & \lstick{$\ket{0}_{n-1}$} & \qw & \qw & \qw \\
        & \lstick{$\ket{0}_c$} & \qw & \qw & \qw
    \end{quantikz}
    \; $\equiv$ \begin{quantikz}[row sep={0.65cm,between origins}]
    & \lstick{$\ket{0}_{(n)}$} & [4mm] \gate[]{\hat{P}} \qwbundle{n} & \gate[wires=2]{\hat{C}_k} & \qw \\
    & \lstick{$\ket{0}_{c}$} & \qw & \qw & \qw \\
    \end{quantikz}
    \caption{The state of play so far: the probability loading and comparator subroutines. Note that any intermediate carry ancilla qubits aren't shown.}
    \label{circ:probloadandCk}
\end{figure}

Given $\hat{C}_k$, the next step is to encode the linear part of the payoff function in the amplitude of a payoff ancilla qubit (our analogous qudit protocol only requires a payoff qubit, and not a qudit). This is done using controlled-$Y$ rotations\footnote{A simple, `uncontrolled' rotation of a qubit $\ket{0}$ about the $Y$ axis of its Bloch sphere gives $\ket{0} \to \cos(\theta)\ket{0} + \sin(\theta)\ket{1}$, where $\theta$ is a fixed angle. However, for a \textit{controlled} rotation, with the qudit register in the current context acting as the control, $\theta$ is not fixed but is generally a function of the register's state. See \cite{nielsen2010quantum} for further details.\label{footnote:rotation}} where, in our case, the asset price qudit register will act as the control. This encoding isn't exact: it is not possible to perfectly encode a linear function in the amplitude of the ancilla using qubit rotations on a machine with finite resources since these rotations only faithfully encode trigonometric functions. However, as outlined in \cite{woerner2019quantum}, we will shift and scale the rotations such that an accurate approximation of the true payoff function is encoded.

Since the $Y$ rotations will approximately encode the payoff function samples \mbox{$f(i) \equiv f(s_i)$} $= \max(0, s_i - K)$ (for the sampled asset values $\{ s_i \}$ in Eq. \ref{eqn:affinemap}), and since quantum amplitude estimation (either with or, in our case, without phase estimation) requires that the rotation angles be restricted such that the mapping from the states $\ket{i}_{(n)}$ to the angles be one-to-one \cite{brassard2002quantum}, we must shift and scale the payoffs $f(i)$ accordingly. First, define

\begin{equation}
\tilde{f}(i) = 2 \, \frac{f(i) - \min_i(f(i))}{\max_i(f(i)) - \min_i(f(i))} - 1 = 2 \, \frac{\max(0, s_i - K)}{s_{d^n - 1} - K} - 1 \in [ -1, 1 ].
\label{eqn:ftilde}
\end{equation}

Next, the thing that is measured in practice for the traditional qubit, as well as the current qudit, option pricing scheme (when amplitude estimation without phase estimation \cite{suzuki2020amplitude} is employed), is the state of the payoff ancilla qubit. Per Born's rule \cite{born1926quantenmechanik}, the probability of measuring either of the two possible computational basis states is contained in the probability amplitudes of this payoff qubit. Since we're encoding the payoff $f$ by performing rotations on its state, the amplitude of measuring the $\ket{1}$ state will be $\sin^2(\theta)$ (c.f. footnote \ref{footnote:rotation}), or, more accurately - since our qudit register is in a superposition of states - a convex combination of the sine-squared function, $\sum_{i=0}^{d^n-1} p_i \sin^2(\theta_i)$, for some $\theta_i$'s. This function, $\sin^2(\theta)$, is one-to-one and onto on the domain $[0, \frac{\pi}{2}]$ and codomain $[0, 1]$. Therefore, we shift and scale $\tilde{f}(i)$ further, to $c\tilde{f}(i) + s$, where $s \equiv \frac{\pi}{4}$ is the domain midpoint and $0 < c \leq \frac{\pi}{4}$ is a small constant such that $c\tilde{f}(i) + s \in [0, 1]$ for all $\tilde{f}(i)$ (more detailed reasons for these choices, as well as an optimal scheme for choosing $c$, can be found in \cite{woerner2019quantum}). As such, $c\tilde{f}(i) + s$ equals either $\frac{\pi}{4} - c$ if $i<k$, or $\frac{2c(s_i - K)}{s_{d^n - 1} - K} + \frac{\pi}{4} - c$ if $i \geq k$.

We are now in a position to outline the second part of the European call payoff loading subroutine:

\begin{itemize}
\item[2.2] Construct an operator $\hat{L}_{f}$, using controlled-$Y$ rotations with the qudit register and comparator qubit as the controls, to encode $f$ in the amplitude of the payoff qubit as

\begin{equation}
    \ket{i}_{(n)} \ket{\text{x}}_c \ket{0}_p \xrightarrow{\hat{L}_f}
    \begin{cases}
    \ket{i}_{(n)} \ket{0}_c \left( \cos(\frac{\pi}{4} - c) \ket{0}_p + \sin(\frac{\pi}{4} - c) \ket{1}_p \right) & \text{if x} = 0, \\
    \ket{i}_{(n)} \ket{1}_c \left( \cos(\frac{2c(s_i - K)}{s_{d^n - 1} - K} + \frac{\pi}{4} - c) \ket{0}_p + \sin(\frac{2c(s_i - K)}{s_{d^n - 1} - K} + \frac{\pi}{4} - c) \ket{1}_p \right) & \text{if x} = 1.
    \end{cases}
\label{eqn:Loperator}
\end{equation}

\end{itemize}

The circuit implementation of $\hat{L}_f$ is explained in detail in Appendix \ref{subsec:payoffencoding}. Finally, recall that we cannot perfectly encode $f$ in the amplitude of the payoff ancilla. Indeed, $\hat{L}_f$ does not encode $f$ in the amplitude, but rather $\sin^2(c\tilde{f}(i) + s)$. However, for small enough $\theta$, note that $\sin^2(\theta + \frac{\pi}{4}) = \theta + \frac{1}{2} + \mathcal{O}(\theta^3)$: $\theta + \frac{1}{2}$ provides an adequate approximation\footnote{We could better encode $c\tilde{f}(i) + s$ in the amplitude of the payoff qubit by modifying $\hat{L}_f$ to include higher-order terms. However, this would involve a deeper circuit with far more gates.}. As such, combining Eqs. \ref{eqn:CPresult} and \ref{eqn:Loperator} and defining $\hat{A} \equiv \hat{L}_f \hat{C}_k \hat{P}$, we get

\begin{align}
\ket{0}_{(n)} \ket{0}_{c} \ket{0}_{p} \xrightarrow{\hat{A}} & \sum_{i < k} \sqrt{p_i} \ket{i}_{(n)} \ket{0}_c \left( \cos(\frac{\pi}{4} - c) \ket{0}_p + \sin(\frac{\pi}{4} - c) \ket{1}_p \right) \nonumber \\
+ & \sum_{i \geq k} \sqrt{p_i} \ket{i}_{(n)} \ket{1}_c \left( \cos(\frac{2c(s_i - K)}{s_{d^n - 1} - K} + \frac{\pi}{4} - c) \ket{0}_p + \sin(\frac{2c(s_i - K)}{s_{d^n - 1} - K} + \frac{\pi}{4} - c) \ket{1}_p \right) \nonumber \\
= & \ket{\Psi_0} \ket{0}_p + \ket{\Psi_1} \ket{1}_p \equiv \ket{\Psi},
\label{eqn:Aresult}
\end{align}

\begin{figure}[h]
    \centering
    \begin{quantikz}[align equals at=2, row sep={0.65cm,between origins}]
    & \lstick{$\ket{0}_{(n)}$} & [4mm] \gate[]{\hat{P}} \qwbundle{n} & \gate[wires=2]{\hat{C}_k} & \gate[wires=3]{\hat{L}_f} & \qw \\
    & \lstick{$\ket{0}_{c}$} & \qw & \qw & \qw & \qw \\
    & \lstick{$\ket{0}_{p}$} & \qw & \qw & \qw & \qw \\
    \end{quantikz}
    \; $\equiv$ \begin{quantikz} [row sep={0.65cm,between origins}]
    & \lstick{$\ket{0}$} & \gate[]{\hat{A}} & \qw
    \end{quantikz}
    \caption{Probability loading $\hat{P}$, comparator $\hat{C}_k$ and payoff encoding $\hat{L}_f$ subroutines.}
    \label{circ:probloadandCkandpayoff}
\end{figure}
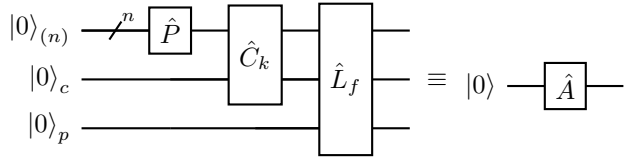

\noindent
(see Fig. \ref{circ:probloadandCkandpayoff}). Finally, after having prepared the composite state $\ket{\Psi}$, the probability of measuring the payoff qubit in the $\ket{1}_p$ state is

\begin{align}
P_1 & = \bra{\Psi_1} \prescript{}{p}{\braket{1}{\Psi}} = \bra{\Psi_1} \prescript{}{p}{\braket{1}{\Psi_1}} \ket{1}_p = \sum_{i < k} p_i \sin^2 \left( \frac{\pi}{4} - c \right) + \sum_{i \geq k} p_i \sin^2 \left( \frac{2c(s_i - K)}{s_{d^n - 1} - K} + \frac{\pi}{4} - c \right) \nonumber \\
& \approx \sum_{i < k} p_i \left( \frac{1}{2} - c \right) + \sum_{i \geq k} p_i \left( \frac{2c(s_i - K)}{s_{d^n - 1} - K} + \frac{1}{2} - c \right) = \frac{1}{2} - c + \frac{2c}{s_{d^n - 1} - K} \, \text{E}_{\text{P}} \left[ f(S_T) \right],
\label{eqn:P1probability}
\end{align}

\noindent
up to third-order terms. Once we have outlined an efficient way of measuring $P_1$, Eq. \ref{eqn:P1probability} gives us a way of recovering the expected payoff $\text{E}_{\text{P}}[f(S_T)]$.

However, how do we measure $P_1$? Na\"{i}vely repeatedly preparing $\ket{\Psi}$ and then measuring the state of the payoff qubit in an effort to create an estimator of $P_1$ provides no quantum advantage: indeed, while doing this would eventually give the correct answer, many measurement samples would be needed as this approach is essentially classical Monte Carlo integration which, as mentioned previously, converges as $\mathcal{O}(M^{-1/2})$ (where $M$ is the number of measurement samples taken).

The real advantage offered by quantum computers in pricing complex derivatives stems ultimately from the amplitude amplification inherent to QAE. In a quantum setting, the notion that is equivalent to the number of classical measurement samples taken, is the number of times the operator, or oracle, $\hat{A}$, is called (we also call this number $M$). As we shall see in the next section, the error in $P_1$ converges as $\mathcal{O}(M^{-1})$, a quadratic speed-up over the equivalent classical algorithm.

\subsection{Step 3: Estimate the probability amplitude}
\label{subsec:QAE}

While a detailed explanation of QAE is given in \cite{brassard2002quantum}, we here outline only the salient points for brevity.

From Eq. \ref{eqn:Aresult}, note that $\hat{A}$, acting on $\ket{0} \equiv \ket{0}_{(n)} \ket{0}_{c} \ket{0}_{p}$, creates a state which can be partitioned into a `good' part, $\ket{\Psi_1}\ket{1}$ (where we have dropped the $p$ subscript on the payoff ancilla), whose amplitude $P_1 = \bra{\Psi_1} \braket{1}{\Psi_1} \ket{1}$ we want, and a `bad' mutually orthogonal part, $\ket{\Psi_0}\ket{0}$. The first step in QAE is to use $\hat{A}$ (see Fig. \ref{circ:probloadandCkandpayoff}) to construct the so-called Grover rotation operator, which is usually denoted by $\hat{Q}$. This operator is a `rotation' in the sense that repeated applications of $\hat{Q}$ to $\ket{\Psi}$ simply rotate the state in the subspace spanned by the vectors $\ket{\Psi_0} \ket{0}$ and $\ket{\Psi_1} \ket{1}$. In other words, $\hat{Q}$ simply changes the relative amplitudes of $\ket{\Psi_0} \ket{0}$ and $\ket{\Psi_1} \ket{1}$ depending on how many times it is applied.

First, note that $\hat{S}_0 := \mathbb{I}_{d^n + 2} - 2 \ketbra{0}{0}$ is an operator which flips the sign of only the ground state and does nothing otherwise\footnote{In reality, any ancilla qubits/qudits, from the comparator subroutines outlined in the Appendix, are also included in the system's composite Hilbert space. These ancillas are just not shown since they are uncomputed in the end.}. Given this, define

\begin{equation}
\hat{S}_{\Psi} := \hat{A} \hat{S}_{0} \hat{A}^{\dagger} = \mathbb{I}_{d^n + 2} - 2 \ketbra{\Psi}{\Psi}, \text{ and } \; \hat{S}_{\Psi_1} := \mathbb{I}_{d^n + 2} - 2 (\mathbb{I}_{d^n + 1} \otimes \ketbra{1}{1}).
\label{eqn:SPsi}
\end{equation}

Here, $\hat{S}_{\Psi}$ represents a reflection about the state $\ket{\Psi}$, whereas $\hat{S}_{\Psi_1}$ has the action of flipping the sign of states in which the payoff ancilla is in the $\ket{1}$ basis state, and does nothing otherwise\footnote{As already noted, creating these operators, and hence $\hat{Q}$, in practice will likely require either decomposing them into whatever fundamental gate set will be available on any potential qudit hardware, or will necessitate feeding these desired operators as input into quantum optimal control algorithms \cite{werschnik2007quantum, koch2022quantum}. While we have given a scheme to decompose $\hat{A}$ into fundamental matrices (and, as it turns out, $\hat{S}_{\Psi_1}$ can simply be written as a single qubit gate), it isn't clear how to decompose $\hat{S}_{0}$. The purpose of this manuscript is not to discuss either general gate compilation nor optimal control in detail but rather simply serves to both outline the steps needed to calculate the theoretical matrices that represent these operators in a qudit context as well as give a tutorial of quantum option pricing.}. These operators are used to define $\hat{Q}$

\begin{equation}
\hat{Q} := -\hat{S}_{\Psi}\hat{S}_{\Psi_1}.
\label{eqn:Qdef}
\end{equation}

Since $\hat{Q}$ is unitary, its two eigenvalues are pure phases (and are functions of $P_1$). See \cite{brassard2002quantum} for further details of this and the following result but, suffice it to say, this fact can be used to easily compute the action of $\hat{Q}^j$ on $\ket{\Psi}$, for any non-negative integer $j$

\begin{align}
\hat{Q}^j\ket{\Psi} & = \hat{Q}^j \hat{A}\ket{0} \nonumber \\
& = \frac{1}{\sqrt{P_1}} \sin \left( \theta_{P_1}(2j+1) \right) \ket{\Psi_1}\ket{1} + \frac{1}{\sqrt{1 - P_1}} \cos \left( \theta_{P_1}(2j+1) \right) \ket{\Psi_0}\ket{0},
\label{eqn:QjonAzero}
\end{align}

\noindent
where $\sin^2(\theta_{P_1}) = P_1$ (with $0 \leq \theta_{P_1} \leq \frac{\pi}{2}$). Finding $\theta_{P_1}$ hence immediately gives $P_1$. Also, note that $2j+1$ is the number of queries to the oracle $\hat{A}$ (once from preparing state $\ket{\Psi}$, and two from each $\hat{Q}$); recall that the number of calls to this oracle is what is used to gauge the efficiency or convergence of an algorithm. And finally, Eq. \ref{eqn:QjonAzero} roughly gives some intuition behind the quadratic scaling arising from the amplification process: $\sin^2 \left( \theta_{P_1}(2j+1) \right)/P_1 \sim (2j + 1)^2$ is approximately the probability of measuring the `good' state for small $\theta_{P_1}$.

Grover's original search paper \cite{grover1996fast} used $\hat{Q}$, in a far simpler context with only one qubit, to enhance the amplitude of the `good' state. This meant that the good state was more likely to be measured in an experiment. The full quantum amplitude amplification and estimation algorithm \cite{brassard2002quantum} goes further by employing quantum phase estimation. This process, in general as well as in the current context, involves applying many controlled versions of Grover's operator $\hat{Q}$ to create various powers of the eigenvalues of $\hat{Q}$ (which, recall, encode $P_1$). These phase eigenvalues combine together, via phase kickback, in the complex amplitudes of a second qubit register. This register, which acts as the aforementioned control, would simply be a register of \textit{qudits} in a full qudit implementation of QAE \cite{cao2011quantum}. Finally, applying an inverse quantum Fourier transform \cite{Coppersmith1994anapproximate, nielsen2010quantum} and measuring the state of the control register gives an integer whose value maps to a very good estimator of $\theta_{P_1}$ with a high probability. See \cite{nielsen2010quantum} for a full explanation of these concepts.

However, phase estimation and the Fourier transform subroutines are themselves the prohibitively computationally expensive bottlenecks in full amplitude estimation algorithms carried out on current, noisy qubit or qudit NISQ-era hardware. As such, various approaches which maintain the quadratic speed-up while forgoing both the phase estimation and Fourier transform subroutines have been proposed \cite{suzuki2020amplitude, aaronson2020quantum, grinko2021iterative}). Arguably the simplest method among these is that outlined in \cite{suzuki2020amplitude}, where maximum likelihood estimation is carried out on measurement data collected after various numbers of ordinary, un-controlled $\hat{Q}$ operations are applied to the quantum system. This is the technique we'll detail and simulate below.

We can now state the final step in the option pricing scheme

\begin{figure}[h]
    \centering
    \begin{quantikz}[align equals at=2, row sep={0.65cm,between origins}]
    & \lstick{$\ket{0}_{(n)}$} & [4mm] \gate[wires=3]{\hat{A}} \qwbundle{n} & \gate[wires=3]{\hat{Q}} & \gate[wires=3]{\hat{Q}} & \qw \ \ldots\ & \gate[wires=3]{\hat{Q}} & \qw \\
    & \lstick{$\ket{0}_{c}$} & \qw & \qw & \qw & \qw \ \ldots\ & \qw & \qw \\
    & \lstick{$\ket{0}_{p}$} & \qw & \qw & \qw & \qw \ \ldots\ & \qw & \meter{} \\
    \end{quantikz}
    \caption{Maximum-likelihood amplitude estimation algorithm using Grover's operator $\hat{Q}$.}
    \label{circ:maxlikelihood}
\end{figure}
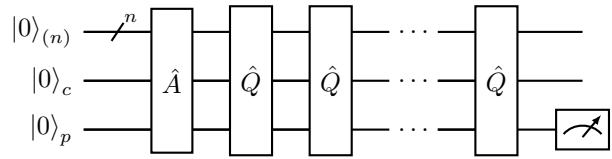

\begin{itemize}
\item[3] Using $\hat{A}$, create the Grover operator $\hat{Q}$ as per Eq. \ref{eqn:Qdef}. Then, apply it $j$ times to the prepared state $\ket{\Psi} = \hat{A}\ket{0}$ and measure the state of the payoff qubit (see Fig. \ref{circ:maxlikelihood}). Repeat this single shot measurement a number of times and for various $j$'s, according to the scheme to be outlined shortly. Finally, use the measurement data to construct a maximum likelihood (ML) function from which the ML estimate of $\theta_{P_1}$ can be gleaned.
\end{itemize}

Maximum likelihood estimation (MLE) is an important technique in modern statistics. Here we give a brief summary of MLE and tailor the ideas to the current option pricing context (see \cite{myung2003tutorial} for an illuminating tutorial on MLE). In general, suppose that one has a statistical model for a random variable and that the model's distribution itself depends on one or more unknown parameters. Furthermore, assume that one can draw samples from the distribution as needed. In this case, MLE provides a technique to deduce the unknown parameters for which, under the chosen mathematical model, the observed sample data is `most likely' to have been observed. This setup can be easily applied to the current context: Eq. \ref{eqn:QjonAzero} represents the probability model for measurements of the random variable (namely the payoff qubit's state) and the unknown parameter we are attempting to deduce is $\theta_{P_1}$.

Following the optimal scheme outlined in \cite{suzuki2020amplitude}, define $m_0 = 0$ and $m_{\ell} = 2^{\ell-1}$, for $\ell = 1, \cdots, T$ for some cutoff $T$\footnote{Note that one could choose a different equation for the $m_{\ell}$'s. However, this optimal exponential sequence asymptotically maintains the quadratic speed-up \cite{suzuki2020amplitude}.}. The $m_{\ell}$'s will correspond to the number of times $\hat{Q}$ is applied to the state $\hat{A}\ket{0}$, i.e. $\hat{Q}^{m_{\ell}} \hat{A} \ket{0}$, in different experimental runs. Furthermore, take $N$ to be the fixed number of single-shot measurements that will be taken of the payoff qubit's state for each run $m_{\ell}$.

Any single measurement of the payoff qubit's state is a Bernoulli trial with $\sin^2 \left(\theta_{P_1}(2 m_{\ell} +1)\right)$ being the probability of measuring the `good' state $\ket{1}$ and $\cos^2 \left(\theta_{P_1}(2 m_{\ell} +1)\right)$ the probability of measuring the `bad' state $\ket{0}$. After performing this single measurement, set $s_{\ell} = 1$ if we measure the good state and $s_{\ell} = 0$ if we measure the bad one. In this case, the likelihood function is

\begin{equation}
\mathcal{L}_{\ell}(\theta_{P_1} | s_{\ell}) = \left[ \sin^2 \left(\theta_{P_1}(2 m_{\ell} +1)\right) \right]^{s_{\ell}} \left[ \cos^2 \left(\theta_{P_1}(2 m_{\ell} +1)\right)\right]^{1-s_{\ell}}.
\label{eqn:onesingleL}
\end{equation}

If we repeat this measurement $N$ times for a fixed $m_{\ell}$ (so that the events are independent and identically distributed) and let $s_{\ell} \in \{0, \cdots, N\}$ denote the number of times the `good' state is measured, this experiment is described by a binomial distribution with the likelihood function

\begin{equation}
\mathcal{L}_{\ell}(\theta_{P_1} | s_{\ell}) = \binom{N}{s_{\ell}} \left[ \sin^2 \left(\theta_{P_1}(2 m_{\ell} +1)\right) \right]^{s_{\ell}} \left[ \cos^2 \left(\theta_{P_1}(2 m_{\ell} +1)\right) \right]^{N-s_{\ell}}.
\label{eqn:Lfixedell}
\end{equation}

Next, even though each $s_{\ell}$ is not identically distributed for different $\ell$'s, the samples are generated independently from the same joint distribution. As such, after repeating the single shot measurement $N$ times for each $m_{\ell}$ and recording each $s_{\ell}$, we can arrive at a single likelihood function for the unknown parameter $\theta_{P_1}$ by simply multiplying the individual $\mathcal{L}_{\ell}$'s together

\begin{equation}
\mathcal{L}(\theta_{P_1} | \{ s_{\ell} \}) = \prod_{\ell=0}^{T} \mathcal{L}_{\ell}(\theta_{P_1} | s_{\ell}).
\label{eqn:likelihoodL}
\end{equation}


In a rough sense, $\mathcal{L}$ represents the probability of observing the set of random variables $\{ s_{\ell} \}$, as a function of $\theta_{P_1}$. The last step of MLE entails `maximising' this probability by finding the value of the parameter which maximises $\mathcal{L}$. This estimate of $\theta_{P_1}$, which we denote $\hat{\theta}_{P_1}$, is hence

\begin{equation}
\hat{\theta}_{P_1} = \arg_{0 \leq \theta_{P_1} \leq \frac{\pi}{2}} \, \max \mathcal{L} \left( \theta_{P_1} | \{ s_{\ell} \} \right).
\label{eqn:thetaestimator}
\end{equation}

$\hat{\theta}_{P_1}$ is the value for which the data samples $\{ s_{\ell} \}$ are `most likely' to have been observed and is generally found in practice using numerical solvers.

We finally have all of the ingredients that are needed to price a European call option on a quantum computer with a qudit register, since the estimate $\hat{\theta}_{P_1}$ is mapped to an estimate for $P_1$, which in turn can be easily classically mapped to the expected payoff via Eq. \ref{eqn:P1probability}. Note that the total number of oracle queries (i.e. the number of times $\hat{A}$ is applied), namely $M$, in this experiment is $M = \sum_{\ell = 0}^T N (2m_{\ell} + 1)$. Although proving so is beyond the scope of the current text, it can be shown that for the amplitude estimation without phase estimation scheme in both a qubit and qudit context, the estimation error $\hat{\epsilon}_{P_1} = \sqrt{\text{E}[(\theta_{P_1} - \hat{\theta}_{P_1})^2]}$ asymptotically scales as $\mathcal{O}(M^{-1})$ (see \cite{suzuki2020amplitude, aaronson2020quantum} for details). As such, the full quantum algorithm here presents a theoretical quadratic speedup compared with typical derivative pricing calculations employing classical Monte Carlo techniques. However, as mentioned before, path-independent options, such as the European call option presented here, are simple enough to unlikely benefit much from quantum derivative pricing. On the other hand, path-dependent financial options, such as the barrier options briefly mentioned at the end of section \ref{sec:classfin} as well as the real-world financial options discussed in \cite{chakrabarti2021threshold}, are likely to benefit the most from advances in quantum computing \cite{chakrabarti2021threshold}. While we close off this manuscript by simulating the entire stack for a relatively simple path-independent option, its purpose wasn't to delve deeply into more complex financial derivatives but rather to both discuss the basics of derivative pricing for those unfamiliar with the subject as well as to outline the quantum computing algorithm behind a simple use case, in enough mathematical detail, so as to easily allow the interested reader to disassemble and then `stitch together' the constituent subroutines of the algorithm when applying quantum computing to more complex derivatives.

\section{European call simulations}
\label{sec:simulations}

We finally simulate the entire European call option pricing scheme using a custom \textsc{Python} package, from sampling the asset's theoretical distribution and constructing the appropriate circuit subroutines, to performing a maximum likelihood estimation on the simulated measurement results. The scheme presented in section \ref{sec:quditalg} above is similar to what a future implementation of financial derivative pricing on a qudit-based quantum computer could potentially entail.

\begin{figure}[h]
\centering
    \begin{subfigure}[b]{0.45\textwidth}
    \centering
    \includegraphics[width=\textwidth]{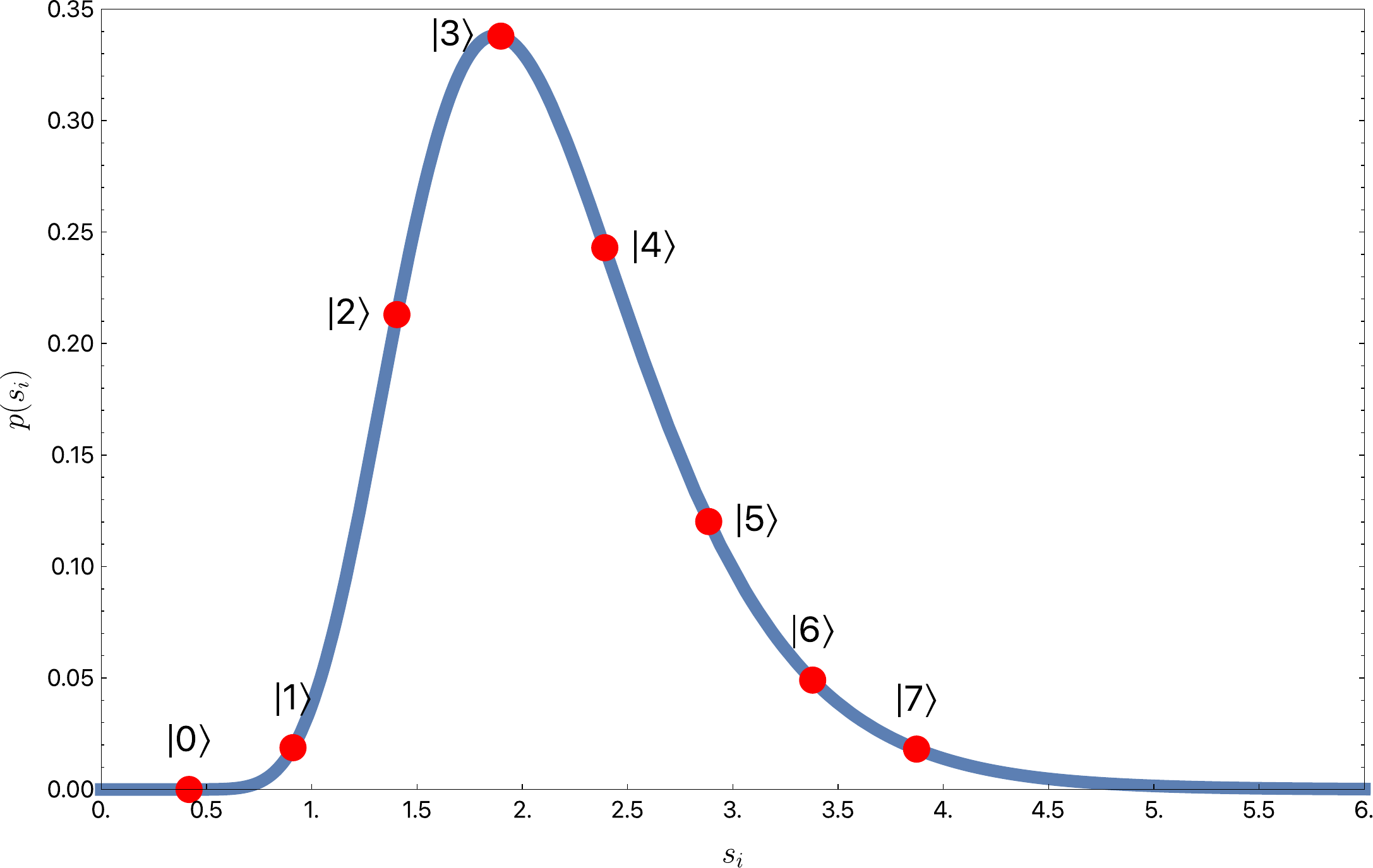}
    \caption{BSM model asset price distribution and sampled points, assuming an $8$-dimensional asset price register.}
    \label{subfig:probgraph1}
    \end{subfigure}
    \hfill
    \begin{subfigure}[b]{0.45\textwidth}
    \centering
    \includegraphics[width=\textwidth]{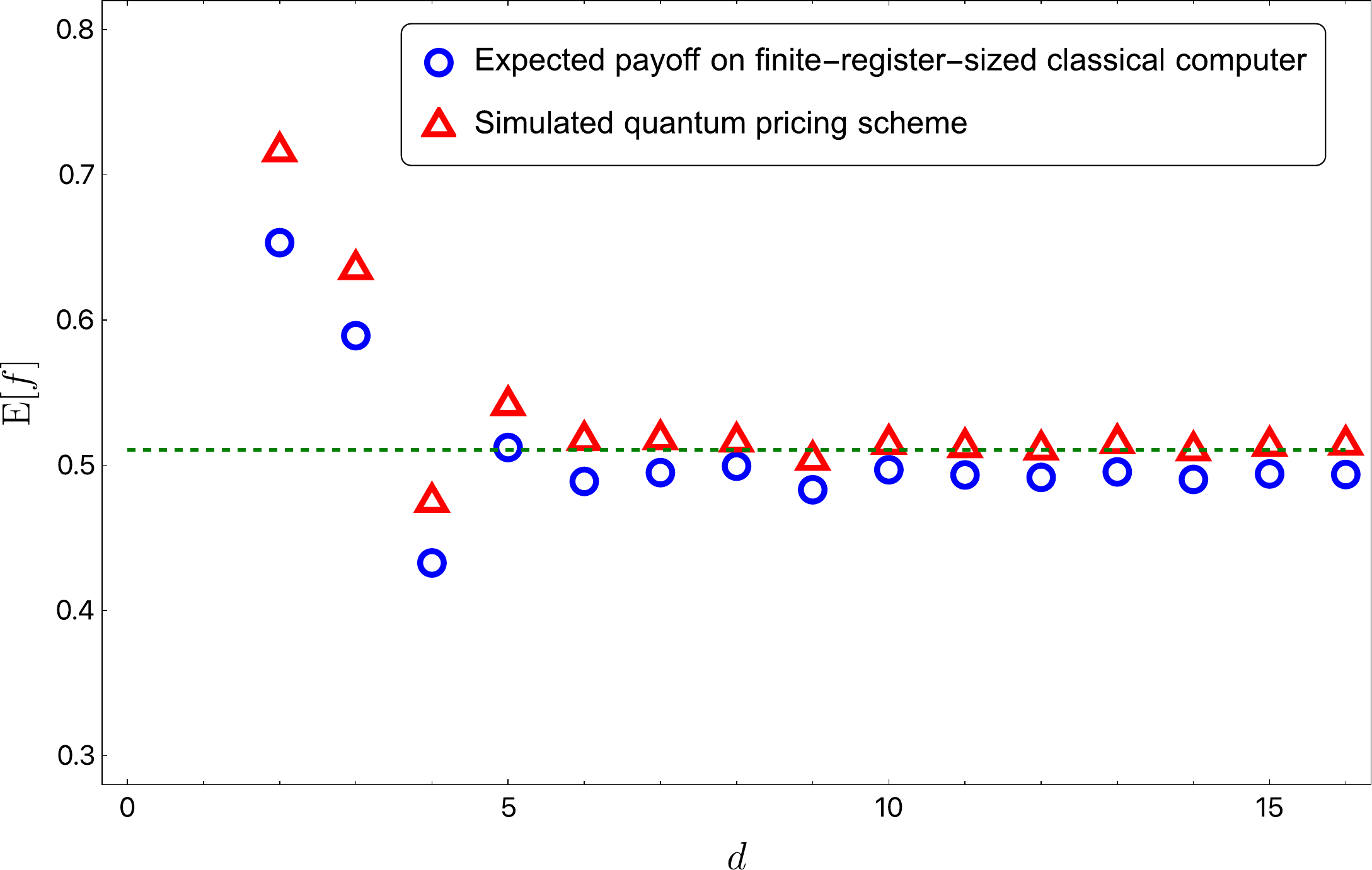}
    \caption{Corresponding expected payoff of European call option vs. total Hilbert space dimension.}
    \label{subfig:results1}
    \end{subfigure}
\caption{Simulation of the expected payoff of a European call option. (a) The encoded distribution, namely Eq .\ref{eqn:probdensSt}, with $S_0 = 2.0, t = 365/365, \alpha = 0.07$ and $\sigma = 0.3$.
(b) The analytic expected payoff (dashed line - assuming a strike price $K$ of $1.7$), the expected payoff given a classical register of size $d$ (blue circles), and the corresponding quantum simulation for a single asset qudit register ($n=1$) with a Hilbert space dimension of $d$ (red triangles). $N=100$ shots are taken for each $m_{\ell}$, and $T=7$ (giving $M=26200$).}
\label{fig:simulation1}
\end{figure}

Fig. \ref{subfig:probgraph1} gives the distribution of the random price of a single asset arising from the Black-Scholes-Merton model, with the chosen parameters in the caption. Since this distribution has relatively flat tails, truncating and discretising the distribution doesn't introduce significant errors. This may not be the case for more volatile or some realistic, market-implied distributions. However, as quantum computers improve over time, such cases could be studied by increasing either the logical qubit/qudit count or increasing the effective Hilbert space dimension of existing qudits. This would allow one to produce a more fine-grained discrete approximation of the true distribution. The latter, namely exploiting the theoretically larger state space of qudit-based hardware, isn't possible when working strictly with qubits and hence presents one possible boon of qudits, all else being equal. However, a trade-off between the experimental time and effort required to increase the dimension of a qudit and the logarithmic scaling of the information that is able to be stored in a qudit, would need to be considered.

Fig. \ref{subfig:results1} gives the corresponding expected payoff as a function of the size of the asset price's qudit register, or equivalently, the number of samples of the underlying probability distribution taken. The discrepancy between both the classical and quantum simulation data points with the analytic payoff is an inevitable result of truncating the domain of the distribution, a process which discards the large but relatively unlikely asset values. For a fixed number of qubits/qudits, increasing the size of the truncated region would capture more extreme prices but with the drawback of producing a more coarse-grained discretisation. However, more important is the fact that the qudit algorithm's expected payoff data points (the red triangles) are close to the payoff calculated on a finite-resource classical computer (blue circles): the former even approaches the latter as the size of the Hilbert space $d$ increases. The classical and quantum payoffs are close to one another after a relatively small increase in Hilbert space size, suggesting that a qudit register of modest size may be sufficient to match the payoff possible on a finite-resource classical device, within error. The remaining small discrepancy between the quantum and classical simulation data points is due to a combination of the constant error arising from encoding the linear payoff function using trigonometric rotations \cite{woerner2019quantum} as well as the $\mathcal{O}(M^{-1})$ error arising from QAE itself. This discrepancy could be decreased further by including higher-order terms in the payoff loading subroutine $\hat{L}_f$ (which would result in a deeper circuit), by perhaps increasing the number of shots $N$ for each unique circuit $\ell$, as well as increasing the number of unique circuits $\ell$ that are measured too. It should be noted, however, that these aforementioned discrepancies arise from the theoretical derivative pricing algorithm itself, to say nothing about the physics- and device-level errors of current NISQ-era hardware. For example, a future quantum compiler (a good explanation of the technical details of which, in the case of \textsc{IBM}'s widely-used qubit compiler, can be found in \cite{cross2019validating}) that would compile the required unitary operations ($\hat{A}$ and $\hat{Q}$ in our case) into hardware-native operations, will be subject to computational and coherent errors. Furthermore, both physical qubit and qudit systems have inherent sources of noise on a fundamental physics level, the study of which is an active area of research \cite{mueller2019towards}.

\begin{figure}[h]
\centering
    \begin{subfigure}[b]{0.45\textwidth}
    \centering
    \includegraphics[width=\textwidth]{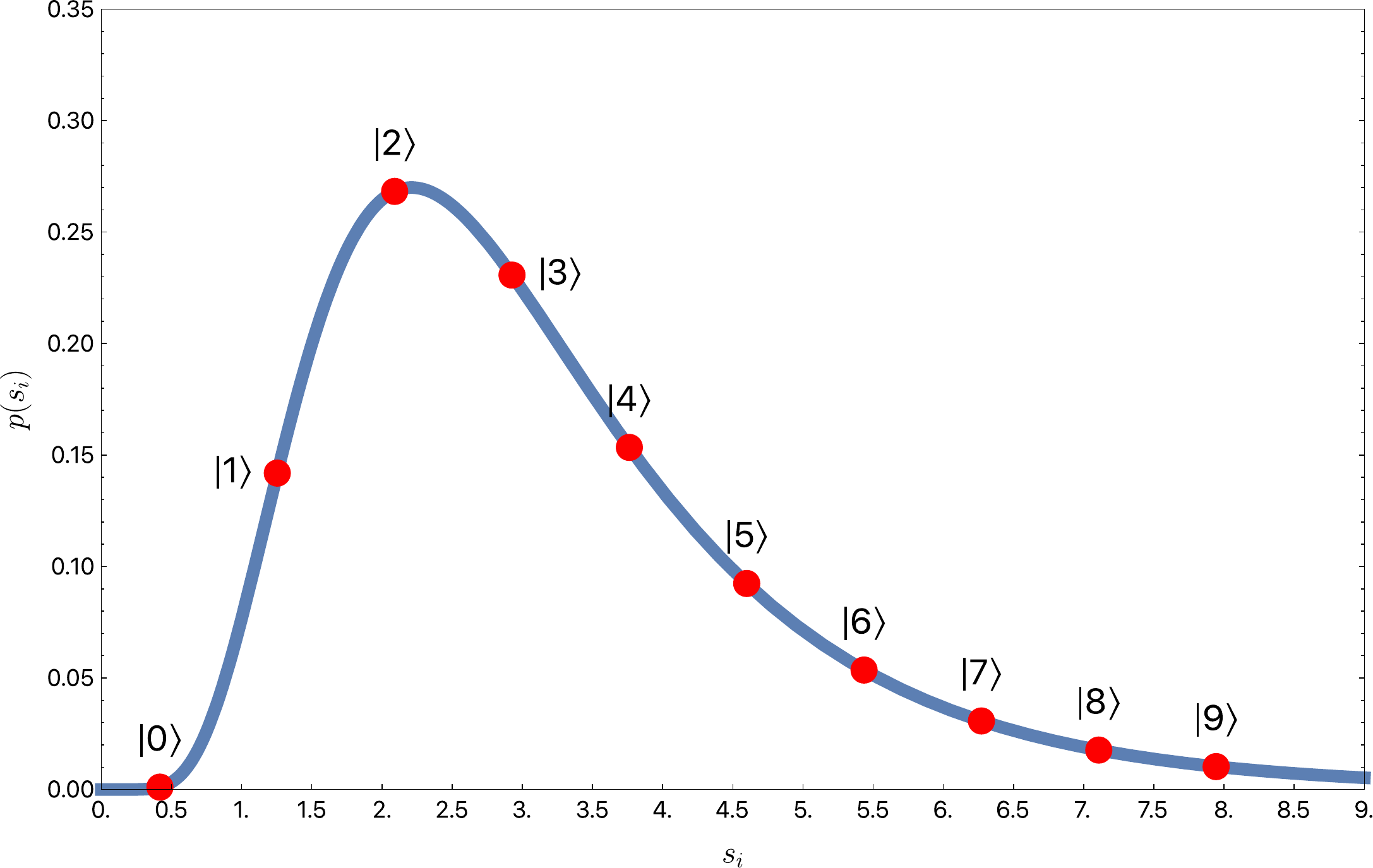}
    \caption{BSM model, with a higher volatility, and a $10$-dimensional asset price register's sample points.}
    \label{subfig:probgraph2}
    \end{subfigure}
    \hfill
    \begin{subfigure}[b]{0.45\textwidth}
    \centering
    \includegraphics[width=\textwidth]{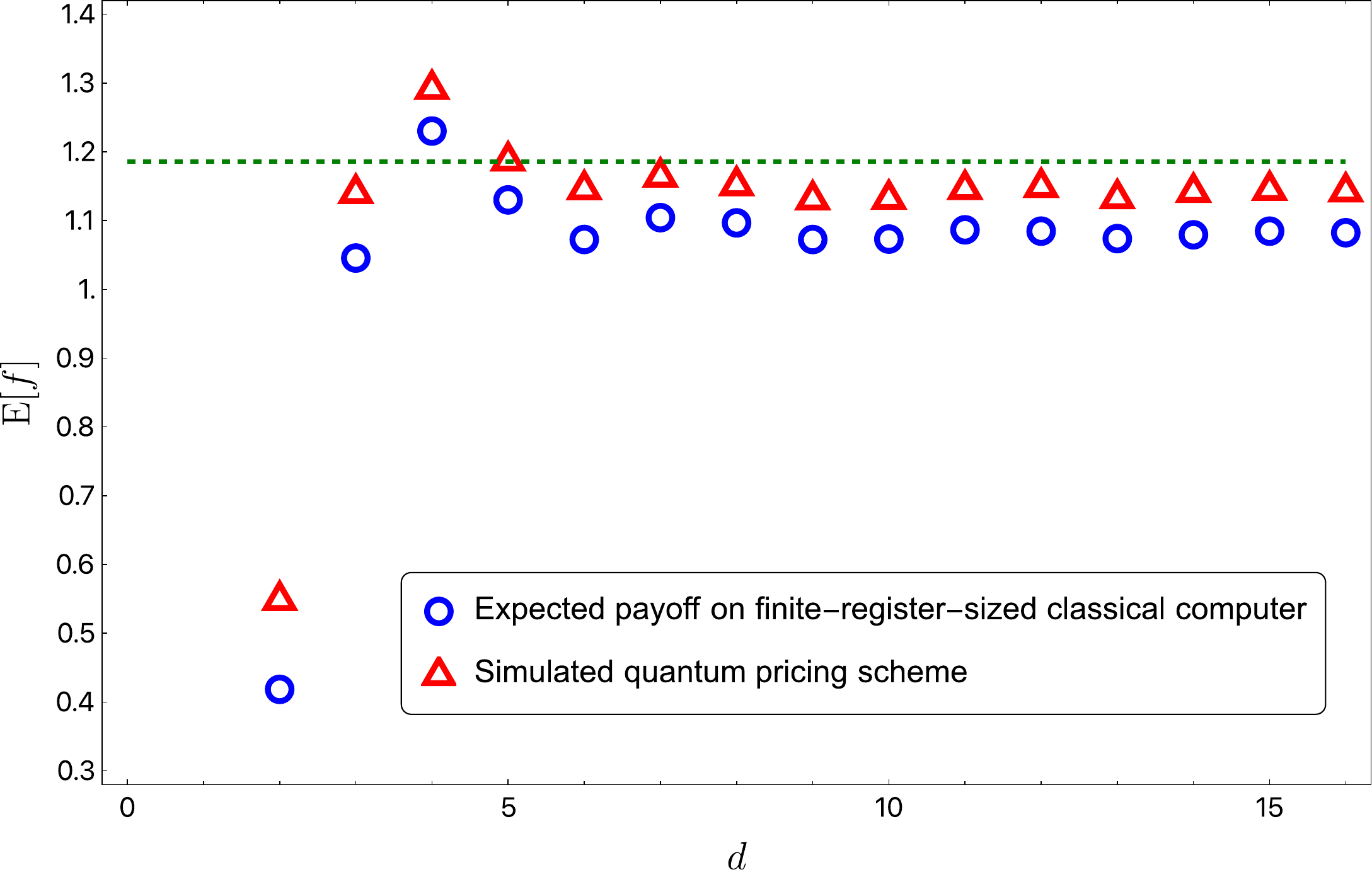}
    \caption{Corresponding expected payoff of option vs. total Hilbert space dimension.}
    \label{subfig:results2}
    \end{subfigure}
\caption{Simulation of a European call option with a more volatile underlying asset. (a) The same analytic distribution and parameters as Fig. \ref{fig:simulation1}, except now with $S_0 = 3.0$ and $\sigma = 0.5$. (b) The corresponding simulated payoff data (with a strike price now of $2.2$), with the dashed line indicating the analytic value of the expected payoff.}
\label{fig:simulation2}
\end{figure}

Fig. \ref{fig:simulation2} gives simulated results, also for a European call option, but one based on a more volatile underlying asset. The larger volatility increases the statistical spread of the distribution which in turn increases the probability mass of extreme asset prices. The simulation in Fig. \ref{subfig:results2}, as was the case in Fig. \ref{subfig:results1}, truncates the asset price to within three standard deviations of the mean. The larger spread of this distribution means that extreme asset prices are not as well accounted for in the probability loading subroutine $\hat{P}$. As a result, there is a larger discrepancy in Fig. \ref{subfig:results2} between the analytic payoff and both the classical finite-resource and quantum payoffs, as expected. However, as before, the classical finite-resource and quantum payoff simulations again match, up to the expected theoretical error.

\section{Conclusion}
\label{sec:conclusion}

We have presented a tutorial on the problem of European option pricing and the algorithm steps one might go through when realistically pricing such an option on a future quantum computer with qudit information carriers. While the primer on derivatives in section \ref{sec:classfin} is admittedly brief and doesn't do the enormous field of financial mathematics much justice, we have endeavoured to both provide the necessary basic details for those unfamiliar with the theory behind financial options as well as outline the BSM model, which is the starting point both in the historical development of financial mathematics as well as a natural point from which the interested reader may begin exploring the topic. It must again be emphasised that the simple European options discussed in section \ref{sec:classfin} can be efficiently priced classically; more complex, realistic, path-dependent financial options are those which stand to benefit the most from quantum computers (see \cite{chakrabarti2021threshold} for an excellent explanation of real-world derivatives as well as resource estimates for their pricing on a qubit-based quantum computer). However, given that quantum computing hardware is still in its relative infancy, comprised of relatively small numbers of noisy qubits or qudits, studying the simplest financial use cases will remain the benchmark for some time.

Armed with this basic understanding, section \ref{sec:quditalg} described the probability loading, payoff computation and amplitude estimation routines in quantum derivative pricing. While loading a log-normal distribution - of which the European call's distribution is an example - into a register of qudits can be done efficiently, detailed comments on the ease with which more general distributions could be encoded can only be made in light of knowledge of the native operations a potential qudit-based quantum computer will have access to. For example, for microwave cavities coupled to superconducting qubits (\cite{haroche2020from, blais2021circuit}), gate sets comprised of SNAP plus displacement operations \cite{krastanov2015universal, fosel2020efficient}, or echo-conditional displacement gates plus single qubit rotations \cite{eickbusch2022fast}, provide universal control over their system's Hilbert space. However, these gates differ from those of, say, traditional 2D superconducting qubit chip systems. Regardless, given a universal gate set, it is theoretically possible to encode any arbitrary probability distribution as long as the system is coherent enough to accommodate a deep enough circuit and/or has enough information carriers. All else being equal however, a register of higher-dimensional logical qudits can encode an analytic or market-implied distribution better than a register of logical qubits consisting of an equal number of information carriers. This is important since, as discussed, better distribution encoding is essential in order to better capture outlying events, which tend to have an outsized effect on market dynamics.

In section \ref{subsec:payoff} and Appendix \ref{subsec:comparator}, we discussed a couple of natural qudit analogues of the ordinary qubit comparator: one choice uses fewer gates but requires more ancillas, whereas the other uses only one ancilla but requires a far deeper circuit. Experimental hardware considerations, such as device coherence times and qudit/qubit connectivity, would best dictate which to use in practice. Furthermore, while the ancilla carry qubits in the comparator subroutine as well as the comparator and payoff qubits themselves were indeed qubits, they could very well have been qudits. Ancilla qudits, however, would not provide any advantage over qubits in the option pricing scheme discussed here since these ancillas are merely required to carry one bit of information. However, given that many proposed cavity QED schemes use traditional superconducting qubits as non-linear elements, it is not unreasonable to suppose that a future qudit-based quantum computer may in fact be hybrid in nature, employing both qubits and qudits.

Using this comparator, sections \ref{subsec:payoff}, and \ref{subsec:QAE} detail how the European call payoff is both encoded in the amplitude of a qubit and the expected payoff subsequently recovered using amplitude estimation without phase estimation. The latter, which uses statistical maximum likelihood estimation to forgo phase estimation, will be required in the current NISQ era of quantum computing as a full phase estimation subroutine \cite{kitaev1995quantum, nielsen2010quantum} increases both the depth and width of the qudit circuit beyond near-term capabilities. The scheme that we've discussed is constructed in such a manner so as to reduce practical resource requirements while maintaining the speedup quantum option pricing offers over the usual method of classically pricing an option using Monte Carlo integration.

The subroutines outlined in section \ref{sec:quditalg} also provide many of the ingredients required to construct the qudit algorithms needed to price other types of options. With the addition of further qudit registers to hold multiple assets' prices as well as subroutines which sum and average the integers encoded in these registers, it is possible to use a qudit-based quantum computer to price, for example, the Asian options and barrier options discussed in section \ref{sec:classfin}. Both of these are path-dependent options and contain structures similar to those appearing in other modern, exotic options \cite{colin2021what}. Section \ref{sec:simulations} finally provides a couple of simulations of the entire pricing scheme for the European call option, with different parameters. While the focus of this manuscript has been far more on the journey towards understanding the detailed intersection between quantum computing and derivative pricing and not this simulation `destination', it is important to note that a logical qudit register of only modest size is capable of pricing a European call option with comparable accuracy to a finite-resource classical computer of similar register size. In the current NISQ era, the logarithmic scaling potential of logical qudits still provide a worthwhile advantage over qubits. Furthermore, qudits with correspondingly modest Hilbert space dimensions of the order of $1-10$ are currently being actively studied in existing systems .

Finally, it is worth mentioning that qudits and higher-dimensional Hilbert spaces in general offer other potential advantages. For example, it is possible to encode information in robust continuous-variable states, such as cat states \cite{gerry1997quantum}. This is an example of bosonic encoding \cite{ma2021quantum} whereby the full Hilbert space of electromagnetic signals (in microwave or SRF cavities \cite{alam2022quantum}, for example) are exploited in quantum information encoding and processing, in contrast to constraining oneself to the first two states for each qubit in a superconducting 2D chip. Bosonic encoding schemes allow for novel quantum error mitigation schemes and correction protocols \cite{li2017cat, cai2021bosonic}, an important step on the road to creating a fault-tolerant quantum computer beyond the NISQ era. And beyond quantum computing, into the broader quantum technology sphere, quantum communication channels employing qudits possess larger information bandwidths as well as potentially greater robustness against quantum state cloning attacks \cite{cozzolino2019high}.

\section{Acknowledgements}
\label{sec:acknowledgements}

The author thanks Andy C. Y. Li for his helpful comments. This material is based upon work supported by the U.S. Department of Energy, Office of Science, National Quantum Information Science, Research Centers, Superconducting Quantum Materials and Systems Center (SQMS) under contract number DE-AC02-07CH11359.

\bibliographystyle{ieeetr}
\bibliography{References}

\begin{thebibliography}{10}

\bibitem{ammendola2000devil}
G.~Ammendola, {\em Devil Take the Hindmost: A History of Financial
  Speculation}.
\newblock Taylor \& Francis, 2000.

\bibitem{bsmodel}
F.~Black and M.~Scholes, ``The pricing of options and corporate liabilities,''
  {\em Journal of Political Economy}, vol.~81, no.~3, pp.~637--654, 1973.

\bibitem{baz2004financial}
J.~Baz and G.~Chacko, {\em Financial derivatives: pricing, applications, and
  mathematics}.
\newblock Cambridge University Press, 2004.

\bibitem{hull2003options}
J.~C. Hull, {\em Options futures and other derivatives}.
\newblock Pearson Education India, 2003.

\bibitem{fokker1914mittlere}
A.~D. Fokker, ``Die mittlere energie rotierender elektrischer {D}ipole im
  {S}trahlungsfeld,'' {\em Annalen der Physik}, vol.~348, no.~5, pp.~810--820,
  1914.

\bibitem{planck1917satz}
M.~Planck, ``{\"U}ber einen {S}atz der statistischen {D}ynamik und seine
  {E}rweiterung in der {Q}uantentheorie,'' {\em Sitzungsberichte der
  Preussischen Akademie der Wissenschaften zu Berlin}, vol.~24,
  pp.~324–--341, 1917.

\bibitem{Risken1996}
H.~Risken, {\em Fokker-Planck Equation}, pp.~63--95.
\newblock Berlin, Heidelberg: Springer Berlin Heidelberg, 1996.

\bibitem{investorirrationality}
A.~Damodaran, ``Investor irrationality.''
  \url{https://pages.stern.nyu.edu/~adamodar/New_Home_Page/invfables/investorirrationality.htm}.
\newblock Accessed: 2022-12-25.

\bibitem{roussas2014introduction}
G.~G. Roussas, {\em An introduction to measure-theoretic probability}.
\newblock Academic Press, 2014.

\bibitem{PhysRevA.98.022321}
P.~Rebentrost, B.~Gupt, and T.~R. Bromley, ``Quantum computational finance:
  Monte carlo pricing of financial derivatives,'' {\em Phys. Rev. A}, vol.~98,
  p.~022321, Aug 2018.

\bibitem{chakrabarti2021threshold}
S.~Chakrabarti, R.~Krishnakumar, G.~Mazzola, N.~Stamatopoulos, S.~Woerner, and
  W.~J. Zeng, ``A threshold for quantum advantage in derivative pricing,'' {\em
  Quantum}, vol.~5, p.~463, 2021.

\bibitem{caflisch1998monte}
R.~E. Caflisch, ``Monte carlo and quasi-monte carlo methods,'' {\em Acta
  numerica}, vol.~7, pp.~1--49, 1998.

\bibitem{wang2020qudits}
Y.~Wang, Z.~Hu, B.~C. Sanders, and S.~Kais, ``Qudits and high-dimensional
  quantum computing,'' {\em Frontiers in Physics}, vol.~8, p.~589504, 2020.

\bibitem{chi2022programmable}
Y.~Chi, J.~Huang, Z.~Zhang, J.~Mao, Z.~Zhou, X.~Chen, C.~Zhai, J.~Bao, T.~Dai,
  H.~Yuan, {\em et~al.}, ``A programmable qudit-based quantum processor,'' {\em
  Nature communications}, vol.~13, no.~1, p.~1166, 2022.

\bibitem{montanaro2015quantum}
A.~Montanaro, ``Quantum speedup of monte carlo methods,'' {\em Proceedings of
  the Royal Society A: Mathematical, Physical and Engineering Sciences},
  vol.~471, no.~2181, p.~20150301, 2015.

\bibitem{brassard2002quantum}
G.~Brassard, P.~Hoyer, M.~Mosca, and A.~Tapp, ``Quantum amplitude amplification
  and estimation,'' {\em Contemporary Mathematics}, vol.~305, pp.~53--74, 2002.

\bibitem{kitaev1995quantum}
A.~Y. Kitaev, ``Quantum measurements and the abelian stabilizer problem,'' {\em
  Electron. Colloquium Comput. Complex.}, vol.~TR96, 1995.

\bibitem{nielsen2010quantum}
M.~A. Nielsen and I.~L. Chuang, {\em Quantum Computation and Quantum
  Information: 10th Anniversary Edition}.
\newblock Cambridge University Press, 2010.

\bibitem{suzuki2020amplitude}
Y.~Suzuki, S.~Uno, R.~Raymond, T.~Tanaka, T.~Onodera, and N.~Yamamoto,
  ``Amplitude estimation without phase estimation,'' {\em Quantum Information
  Processing}, vol.~19, pp.~1--17, 2020.

\bibitem{alam2022quantum}
M.~S. Alam, S.~Belomestnykh, N.~Bornman, G.~Cancelo, Y.-C. Chao, M.~Checchin,
  V.~S. Dinh, A.~Grassellino, E.~J. Gustafson, R.~Harnik, {\em et~al.},
  ``Quantum computing hardware for {HEP} algorithms and sensing,'' {\em arXiv
  preprint arXiv:2204.08605}, 2022.

\bibitem{taleb2007black}
N.~N. Taleb, {\em The Black Swan: The Impact of the Highly Improbable}, vol.~2.
\newblock Random House, 2007.

\bibitem{Householder}
M.~Taboga, ``Householder matrix,'' 2021.
\newblock \url{https://www.statlect.com/matrix-algebra/Householder-matrix}.
  Accessed: 2022-12-08.

\bibitem{plesch2011quantum}
M.~Plesch and {\v{C}}.~Brukner, ``Quantum-state preparation with universal gate
  decompositions,'' {\em Phys. Rev. A}, vol.~83, no.~3, p.~032302, 2011.

\bibitem{mottonen2004transformation}
M.~M\"{o}tt\"{o}nen, J.~J. Vartiainen, V.~Bergholm, and M.~M. Salomaa,
  ``Transformation of quantum states using uniformly controlled rotations,''
  {\em Quantum Info. Comput.}, vol.~5, p.~467–473, sep 2005.

\bibitem{vartiainen2004efficient}
J.~J. Vartiainen, M.~M{\"o}tt{\"o}nen, and M.~M. Salomaa, ``Efficient
  decomposition of quantum gates,'' {\em Physical Review Letters}, vol.~92,
  no.~17, p.~177902, 2004.

\bibitem{mottonen2004quantum}
M.~M{\"o}tt{\"o}nen, J.~J. Vartiainen, V.~Bergholm, and M.~M. Salomaa,
  ``Quantum circuits for general multiqubit gates,'' {\em Phys. Rev. Lett.},
  vol.~93, no.~13, p.~130502, 2004.

\bibitem{preskill2018quantum}
J.~Preskill, ``Quantum computing in the {NISQ} era and beyond,'' {\em Quantum},
  vol.~2, p.~79, 2018.

\bibitem{luo2014universal}
M.~Luo and X.~Wang, ``Universal quantum computation with qudits,'' {\em Science
  China Physics, Mechanics \& Astronomy}, vol.~57, pp.~1712--1717, 2014.

\bibitem{grover2002creating}
L.~Grover and T.~Rudolph, ``Creating superpositions that correspond to
  efficiently integrable probability distributions,'' {\em arXiv preprint
  quant-ph/0208112}, 2002.

\bibitem{dixon2020machine}
P.~B. Matthew F.~Dixon, Igor~Halperin, {\em Machine Learning in Finance: From
  Theory to Practice}.
\newblock Springer Cham, 2020.

\bibitem{liu2019pricing}
S.~Liu, C.~W. Oosterlee, and S.~M. Bohte, ``Pricing options and computing
  implied volatilities using neural networks,'' {\em Risks}, vol.~7, no.~1,
  p.~16, 2019.

\bibitem{horvath2021deep}
B.~Horvath, A.~Muguruza, and M.~Tomas, ``Deep learning volatility: a deep
  neural network perspective on pricing and calibration in (rough) volatility
  models,'' {\em Quantitative Finance}, vol.~21, no.~1, pp.~11--27, 2021.

\bibitem{zoufal2019quantum}
C.~Zoufal, A.~Lucchi, and S.~Woerner, ``Quantum generative adversarial networks
  for learning and loading random distributions,'' {\em npj Quantum
  Information}, vol.~5, no.~1, p.~103, 2019.

\bibitem{goodfellow2014generative}
I.~Goodfellow, J.~Pouget-Abadie, M.~Mirza, B.~Xu, D.~Warde-Farley, S.~Ozair,
  A.~Courville, and Y.~Bengio, ``Generative adversarial nets,'' in {\em
  Advances in Neural Information Processing Systems} (Z.~Ghahramani,
  M.~Welling, C.~Cortes, N.~Lawrence, and K.~Weinberger, eds.), vol.~27, Curran
  Associates, Inc., 2014.

\bibitem{Stamatopoulos2020optionpricingusing}
N.~Stamatopoulos, D.~J. Egger, Y.~Sun, C.~Zoufal, R.~Iten, N.~Shen, and
  S.~Woerner, ``Option {P}ricing using {Q}uantum {C}omputers,'' {\em
  {Quantum}}, vol.~4, p.~291, July 2020.

\bibitem{kielpinski2002architecture}
D.~Kielpinski, C.~Monroe, and D.~J. Wineland, ``Architecture for a large-scale
  ion-trap quantum computer,'' {\em Nature}, vol.~417, no.~6890, pp.~709--711,
  2002.

\bibitem{knill2001scheme}
E.~Knill, R.~Laflamme, and G.~J. Milburn, ``A scheme for efficient quantum
  computation with linear optics,'' {\em Nature}, vol.~409, no.~6816,
  pp.~46--52, 2001.

\bibitem{kok2007linear}
P.~Kok, W.~J. Munro, K.~Nemoto, T.~C. Ralph, J.~P. Dowling, and G.~J. Milburn,
  ``Linear optical quantum computing with photonic qubits,'' {\em Reviews of
  Modern Physics}, vol.~79, no.~1, p.~135, 2007.

\bibitem{blais2021circuit}
A.~Blais, A.~L. Grimsmo, S.~M. Girvin, and A.~Wallraff, ``Circuit quantum
  electrodynamics,'' {\em Reviews of Modern Physics}, vol.~93, p.~025005, May
  2021.

\bibitem{haroche2020from}
S.~{Haroche}, M.~{Brune}, and J.~M. {Raimond}, ``From cavity to circuit quantum
  electrodynamics,'' {\em Nature Physics}, vol.~16, no.~3, pp.~243--246, 2020.

\bibitem{romanenko2020three}
A.~Romanenko, R.~Pilipenko, S.~Zorzetti, D.~Frolov, M.~Awida, S.~Belomestnykh,
  S.~Posen, and A.~Grassellino, ``Three-dimensional superconducting resonators
  at {$T \leq 20$} mk with photon lifetimes up to {$\tau=2$} s,'' {\em Phys.
  Rev. Appl.}, vol.~13, p.~034032, Mar 2020.

\bibitem{fosel2020efficient}
T.~Fosel, S.~Krastanov, F.~Marquardt, and L.~Jiang, ``Efficient cavity control
  with {SNAP} gates,'' {\em Bulletin of the American Physical Society}, 2020.

\bibitem{rajeev2023suppressing}
{Google Quantum AI}, ``Suppressing quantum errors by scaling a surface code
  logical qubit,'' {\em Nature}, vol.~614, no.~7949, pp.~676--681, 2023.

\bibitem{krastanov2015universal}
S.~Krastanov, V.~V. Albert, C.~Shen, C.-L. Zou, R.~W. Heeres, B.~Vlastakis,
  R.~J. Schoelkopf, and L.~Jiang, ``Universal control of an oscillator with
  dispersive coupling to a qubit,'' {\em Phys. Rev. A}, vol.~92, p.~040303, Oct
  2015.

\bibitem{cuccaro2004new}
S.~A. Cuccaro, T.~G. Draper, S.~A. Kutin, and D.~P. Moulton, ``A new quantum
  ripple-carry addition circuit,'' {\em arXiv preprint quant-ph/0410184}, 2004.

\bibitem{koren2018computer}
I.~Koren, {\em Computer arithmetic algorithms}.
\newblock AK Peters/CRC Press, 2018.

\bibitem{orts2020review}
F.~Orts, G.~Ortega, E.~Combarro, and E.~Garzón, ``A review on reversible
  quantum adders,'' {\em Journal of Network and Computer Applications},
  vol.~170, p.~102810, 2020.

\bibitem{draper2000addition}
T.~G. Draper, ``Addition on a quantum computer,'' {\em arXiv preprint
  quant-ph/0008033}, 2000.

\bibitem{woerner2019quantum}
S.~Woerner and D.~J. Egger, ``Quantum risk analysis,'' {\em npj Quantum
  Information}, vol.~5, no.~1, p.~15, 2019.

\bibitem{born1926quantenmechanik}
M.~Born, ``Zur quantenmechanik der {S}to{\ss}vorg{\"a}nge,'' {\em Zeitschrift
  f{\"u}r {P}hysik}, vol.~37, pp.~863--867, 1926.

\bibitem{werschnik2007quantum}
J.~Werschnik and E.~Gross, ``Quantum {O}ptimal {C}ontrol {T}heory,'' {\em
  Journal of Physics B: Atomic, Molecular and Optical Physics}, vol.~40,
  no.~18, p.~R175, 2007.

\bibitem{koch2022quantum}
C.~P. Koch, U.~Boscain, T.~Calarco, G.~Dirr, S.~Filipp, S.~J. Glaser,
  R.~Kosloff, S.~Montangero, T.~Schulte-Herbr{\"u}ggen, D.~Sugny, {\em et~al.},
  ``Quantum optimal control in quantum technologies. strategic report on
  current status, visions and goals for research in {E}urope,'' {\em EPJ
  Quantum Technology}, vol.~9, no.~1, p.~19, 2022.

\bibitem{grover1996fast}
L.~K. Grover, ``A fast quantum mechanical algorithm for database search,'' in
  {\em Proceedings of the Twenty-Eighth Annual ACM Symposium on Theory of
  Computing}, STOC '96, (New York, NY, USA), p.~212–219, Association for
  Computing Machinery, 1996.

\bibitem{cao2011quantum}
Y.~Cao, S.-G. Peng, C.~Zheng, and G.-L. Long, ``Quantum {F}ourier {T}ransform
  and {P}hase {E}stimation in {Q}udit {S}ystem,'' {\em Communications in
  Theoretical Physics}, vol.~55, no.~5, p.~790, 2011.

\bibitem{Coppersmith1994anapproximate}
D.~Coppersmith, ``An approximate {F}ourier transform useful in quantum
  factoring,'' in {\em IBM: Research Report RC19642}, 1994.

\bibitem{aaronson2020quantum}
S.~Aaronson and P.~Rall, ``Quantum {A}pproximate {C}ounting, {S}implified,'' in
  {\em Symposium on Simplicity in Algorithms}, pp.~24--32, SIAM, 2020.

\bibitem{grinko2021iterative}
D.~Grinko, J.~Gacon, C.~Zoufal, and S.~Woerner, ``Iterative quantum amplitude
  estimation,'' {\em npj Quantum Information}, vol.~7, no.~1, p.~52, 2021.

\bibitem{myung2003tutorial}
I.~J. Myung, ``Tutorial on maximum likelihood estimation,'' {\em Journal of
  Mathematical Psychology}, vol.~47, no.~1, pp.~90--100, 2003.

\bibitem{cross2019validating}
A.~W. Cross, L.~S. Bishop, S.~Sheldon, P.~D. Nation, and J.~M. Gambetta,
  ``Validating quantum computers using randomized model circuits,'' {\em Phys.
  Rev. A}, vol.~100, p.~032328, Sep 2019.

\bibitem{mueller2019towards}
C.~M{ü}ller, J.~H. Cole, and J.~Lisenfeld, ``Towards understanding
  two-level-systems in amorphous solids: insights from quantum circuits,'' {\em
  Reports on Progress in Physics}, vol.~82, p.~124501, Oct 2019.

\bibitem{eickbusch2022fast}
A.~Eickbusch, V.~Sivak, A.~Z. Ding, S.~S. Elder, S.~R. Jha, J.~Venkatraman,
  B.~Royer, S.~Girvin, R.~J. Schoelkopf, and M.~H. Devoret, ``Fast universal
  control of an oscillator with weak dispersive coupling to a qubit,'' {\em
  Nature Physics}, vol.~18, no.~12, pp.~1464--1469, 2022.

\bibitem{colin2021what}
C.~Dodds, ``What {A}re {E}xotic {O}ptions? 11 {T}ypes of {E}xotic {O}ptions,''
  2021.
\newblock \url{https://www.sofi.com/learn/content/exotic-options/}. Accessed:
  2023-10-23.

\bibitem{gerry1997quantum}
C.~Gerry and P.~Knight, ``Quantum superpositions and {S}chr{\"o}dinger cat
  states in quantum optics,'' {\em American Journal of Physics}, vol.~65,
  no.~10, pp.~964--974, 1997.

\bibitem{ma2021quantum}
W.-L. Ma, S.~Puri, R.~J. Schoelkopf, M.~H. Devoret, S.~M. Girvin, and L.~Jiang,
  ``Quantum control of bosonic modes with superconducting circuits,'' {\em
  Science Bulletin}, vol.~66, no.~17, pp.~1789--1805, 2021.

\bibitem{li2017cat}
L.~Li, C.-L. Zou, V.~V. Albert, S.~Muralidharan, S.~Girvin, and L.~Jiang, ``Cat
  {C}odes with {O}ptimal {D}ecoherence {S}uppression for a {L}ossy {B}osonic
  {C}hannel,'' {\em Physical Review Letters}, vol.~119, no.~3, p.~030502, 2017.

\bibitem{cai2021bosonic}
W.~Cai, Y.~Ma, W.~Wang, C.-L. Zou, and L.~Sun, ``Bosonic quantum error
  correction codes in superconducting quantum circuits,'' {\em Fundamental
  Research}, vol.~1, no.~1, pp.~50--67, 2021.

\bibitem{cozzolino2019high}
D.~Cozzolino, B.~Da~Lio, D.~Bacco, and L.~K. Oxenl{\o}we, ``High-dimensional
  {Q}uantum {C}ommunication: {B}enefits, {P}rogress, and {F}uture
  {C}hallenges,'' {\em Advanced Quantum Technologies}, vol.~2, no.~12,
  p.~1900038, 2019.

\bibitem{pinter2010book}
C.~C. Pinter, {\em A {B}ook of {A}bstract {A}lgebra}.
\newblock Dover Publications, 2010.

\end{thebibliography}

\section{Appendix}
\label{sec:appendix}

\subsection{Comparator subroutine}
\label{subsec:comparator}

Given two classical integers $k$ and $i$, expressed as strings of $n$ digits in base-$d$ (e.g. $k = k_{n-1}k_{n-2}\cdots k_1k_0$), the method of complements - used to compute the difference $i-k$ using addition - is as follows:

\begin{itemize}
\item For each digit $k[j] \equiv k_j$, compute the digit's $d-1$ complement and write the result as a string, namely: \mbox{$(d-1-k_{n-1})(d-1-k_{n-2})\cdots (d-1-k_0)$}
\item Add $1$ to this, calling the result $k^c$
\item Add $i$ and $k^c$, digit by digit. The extra leftmost digit of this sum, at index $n$, is either $0$ or $1$: if it's $0$, then $i$ is less than $k$; if it is $1$, then $i$ is bigger than or equal to $k$
\end{itemize}

For example, in base $5$, with four digits ($n=4$), suppose $i=382_{10}=3012_5$ and $k=329_{10}=2304_5$. In this case, $k^c=2141_5$ and hence $i + k^c = 10203_5$. Since the extra digit, at index position $4$, is $1$, this indicates that $i \geq k$ (and the $4$ least significant digits, namely $0203_5$, gives the difference $i-k=53_{10}$ itself).

This is simple enough in a purely classical case, but we're comparing the qudit quantum state $\ket{i}_{(n)}$ with the classical variable $k$. We hence use a qudit version of the quantum ripple-carry full adder circuit for qubits \cite{cuccaro2004new}, tailored to $k$, to perform this sum. In what follows, we assume that our quantum system, which is comprised of qudits along with some qubits, is equipped with any arbitrary single qudit/qubit gate along with two-qudit/qubit controlled gates. Any unitary matrix operation on the overall composite Hilbert space of the system can then be decomposed into these fundamental operations \cite{wang2020qudits}. In this manuscript, the elementary gates used to construct circuits are chosen simply to draw clean analogies with existing gates in the qubit literature as well as to be able to calculate numerical matrix representations of the required unitary operations. As mentioned in section \ref{subsubsec:resest}, in practice, the matrix operation for the current option pricing algorithm would need to be decomposed into a fundamental gate set offered by the particular `qudit' hardware platform that is available; this is, as yet, unknown.

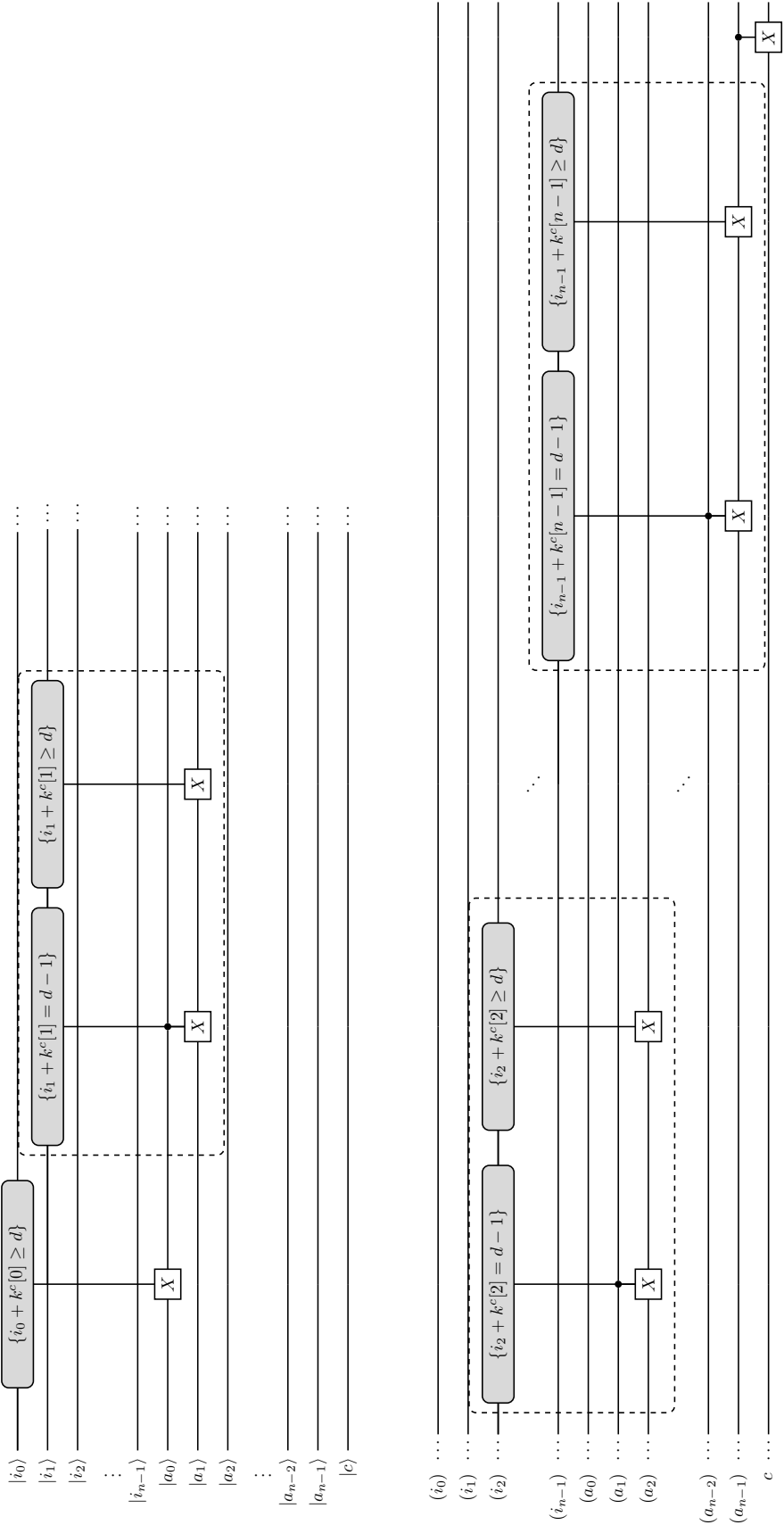
\begin{sidewaysfigure}[ht]
    \centering
    \resizebox{\linewidth}{!}{%
    \begin{quantikz}[row sep={0.6cm,between origins}, column sep=0.4cm]
            & \lstick{$\ket{i_0}$} & \gate[style={rounded corners, fill=gray!30}][0.5cm]{\{ i_0 + k^c[0] \geq d \}} \ctrl{5} & \qw & \qw & \ \dots\ \qw \\
            & \lstick{$\ket{i_1}$} & \qw & \gate[style={rounded corners, fill=gray!30}]{\{ i_1 + k^c[1] = d - 1 \}} \ctrl{5} \gategroup[6,steps=2,style={dashed,rounded corners,inner xsep=2pt},background]{} & \gate[style={rounded corners, fill=gray!30}]{\{ i_1 + k^c[1] \geq d \}} \ctrl{5} & \ \ldots \qw \\
            & \lstick{$\ket{i_2}$} & \qw & \qw & \qw & \ \ldots \qw \\
            & \vdots \hspace{1cm}  & & & & \\
            & \lstick{$\ket{i_{n-1}}$} & \qw & \qw & \qw & \ \ldots\ \qw \\
            & \lstick{$\ket{a_{0}}$} & \gate{X} & \ctrl{1} & \qw & \ \ldots\ \qw \\
            & \lstick{$\ket{a_{1}}$} & \qw & \gate{X} & \gate{X} & \ \ldots\ \qw \\
            & \lstick{$\ket{a_{2}}$} & \qw & \qw & \qw & \ \ldots\ \qw \\
            & \vdots \hspace{1cm} & & & & \\
            & \lstick{$\ket{a_{n-2}}$} & \qw & \qw & \qw & \ \ldots\ \qw \\
            & \lstick{$\ket{a_{n-1}}$} & \qw & \qw & \qw & \ \ldots\ \qw \\
            & \lstick{$\ket{c}$} & \qw & \qw & \qw & \ \ldots\ \qw \\
            & & & & & & \\
            & & & & & & \\
            & \ \ldots\ \lstick{$(i_0)$} & \qw & \qw & \qw & \qw & \qw & \qw & \qw & \qw \\
            & \ \ldots\ \lstick{$(i_1)$} & \qw & \qw & \qw & \qw & \qw & \qw & \qw & \qw \\
            & \ \ldots\ \lstick{$(i_2)$} & \gate[style={rounded corners, fill=gray!30}]{\{ i_2 + k^c[2] = d - 1 \}} \ctrl{5} \gategroup[6,steps=2,style={dashed,rounded corners,inner xsep=2pt},background]{} & \gate[style={rounded corners, fill=gray!30}]{\{ i_2 + k^c[2] \geq d \}} \ctrl{5} & \qw & \qw & \qw & \qw & \qw & \qw \\
            & & & & \ddots & & & & & \\
            & \ \ldots\ \lstick{$(i_{n-1})$} & \qw & \qw & \qw & \gate[style={rounded corners, fill=gray!30}]{\{ i_{n-1} + k^c[n-1] = d - 1 \}} \ctrl{6} \gategroup[7,steps=2,style={dashed,rounded corners,inner xsep=2pt},background]{} & \gate[style={rounded corners, fill=gray!30}]{\{ i_{n-1} + k^c[n-1] \geq d \}} \ctrl{6} & \qw & \qw & \qw \\
            & \ \ldots\ \lstick{$(a_0)$} & \qw & \qw & \qw & \qw & \qw & \qw & \qw & \qw \\
            & \ \ldots\ \lstick{$(a_1)$} & \ctrl{1} & \qw & \qw & \qw & \qw & \qw & \qw & \qw \\
            & \ \ldots\ \lstick{$(a_2)$} & \gate{X} & \gate{X} & \qw & \qw & \qw & \qw & \qw & \qw \\
            & & & & \ddots & & & & & \\
            & \ \ldots\ \lstick{$(a_{n-2})$} & \qw & \qw & \qw & \ctrl{1} & \qw & \qw & \qw & \qw \\
            & \ \ldots\ \lstick{$(a_{n-1})$} & \qw & \qw & \qw & \gate{X} & \gate{X} & \qw & \ctrl{1} & \qw \\
            & \ \ldots\ \lstick{$c$} & \qw & \qw & \qw & \qw & \qw & \qw & \gate{X} & \qw
    \end{quantikz}
    }%
    \caption{Circuit for qudit quantum comparator with $\mathcal{O}(n)$ available ancilla carry qubits. The information carriers denoted with $i$'s are the $d$-dimensional qudits, the $a$'s are the $2$-dimensional ancilla qubits (which we initialise to $\ket{0}$ at the start of the experiment), and qubit $\ket{c}$ is the comparator qubit. Each greyed `control bubble' denotes a set of $i_j$ values (for $j \in \{ 0, \cdots, n-1 \}$) for which the target operation - here only $X$ gates - is applied to the corresponding ancilla qubit $\ket{a_j}$ (since $k$, which is uniquely defined by the strike price $K$, is fixed and known, this circuit is unique and customised for each $k$). For example, the leftmost gate inside the very first dashed box flips the state of $\ket{a_1}$ to $\ket{1}$ only for the basis state $\ket{i_1}$ for which the digit sum of $\ket{i_1}$ and $k^c[1]$ equals $d-1$. In particular, this gate accounts for the case in which the classical addition of $i_1$ and $k^c[1]$ would(not) produce an output carry bit, given a $1$($0$) input carry bit, since the gate itself is also conditioned on the state of the input carry qubit, namely $\ket{a_0}$. The rightmost gate inside the same dashed box accounts for the cases in which $i_1 + k^c[1]$ is large enough to produce an output carry bit of $1$, regardless of the value of the input carry bit (if any control set in this circuit is empty, the entire gate falls away). The presence of an output carry bit is hence encoded in $\ket{a_1}$, which is in turn fed, as an input carry bit, into the gates operating non-trivially on $\ket{i_2}$ and $\ket{a_2}$. This process is repeated in turn for each qudit until the final ancilla, $\ket{a_{n-1}}$, is set to $\ket{0}$ if $i-k$ (equivalently, $i+k^c$) would not produce a carry bit (in which case $i<k$), or $\ket{1}$ if it would. Finally, in order to be able to `re-use' the $a$ qubits in later iterations of $C_k$, the state of $\ket{a_{n-1}}$ is copied to the comparator qubit $\ket{c}$ (also initialised to $\ket{0}$) using one final controlled-$X$ gate, and all of the gates (except the last $CX$) are uncomputed by repeating all of the gates, in reverse order (not shown above). Qubit $\ket{c}$ is the comparator ancilla qubit of step $2.1$ in section \ref{subsec:payoff}.}
\label{circ:type1comparator}
\end{sidewaysfigure}

There are at least two potential approaches to construct circuits for $\hat{C}_k$, depending on how many ancilla carry qubits we wish to include. One approach assumes that we have $\mathcal{O}(n)$ carry qubits available and results in a circuit of depth $\mathcal{O}(n)$ (counted with respect to single-control or double-control qudit/qubit gates). This approach is the qudit analogue of the comparator put to use in \cite{Stamatopoulos2020optionpricingusing}. See Fig. \ref{circ:type1comparator} and its caption for the circuit and detailed explanation thereof (note that any ancilla carry qubits are not shown in the states in the main text of this manuscript).

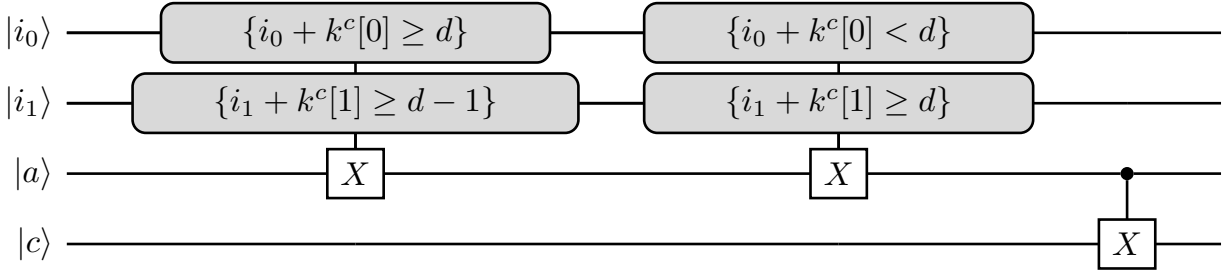
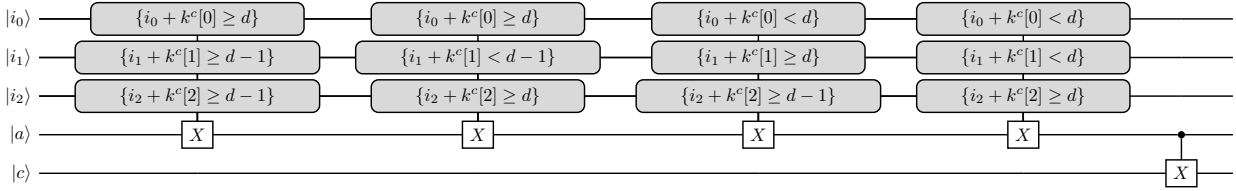
\begin{figure}[h]
    \centering
    \begin{subfigure}[b]{\textwidth}
    \centering
    \resizebox{\linewidth}{!}{%
    \begin{quantikz}[row sep={0.75cm,between origins}, column sep=0.7cm]
            & \lstick{$\ket{i_0}$} & \gate[style={rounded corners, fill=gray!30}]{\{ i_0 + k^c[0] \geq d \}} \ctrl{2} & \gate[style={rounded corners, fill=gray!30}]{\{ i_0 + k^c[0] < d \}} \ctrl{2} & \qw & \qw \\
            & \lstick{$\ket{i_1}$} & \gate[style={rounded corners, fill=gray!30}]{\{ i_1 + k^c[1] \geq d-1 \}} \ctrl{1} & \gate[style={rounded corners, fill=gray!30}]{\{ i_1 + k^c[1] \geq d \}} \ctrl{1} & \qw & \qw \\
            & \lstick{$\ket{a}$} & \gate{X} & \gate{X} & \ctrl{1} & \qw \\
            & \lstick{$\ket{c}$} & \qw & \qw & \gate{X} & \qw
    \end{quantikz}
    }%
    \caption{Two-qudit quantum comparator circuit}
    \label{subcirc:n=2}
    \end{subfigure}%
    \vspace{0.5cm}
    \begin{subfigure}[b]{\textwidth}
    \centering
    \resizebox{\linewidth}{!}{%
    \begin{quantikz}[row sep={0.75cm,between origins}, column sep=0.7cm]
            & \lstick{$\ket{i_0}$} & \gate[style={rounded corners, fill=gray!30}]{\{ i_0 + k^c[0] \geq d \}} \ctrl{3} & \gate[style={rounded corners, fill=gray!30}]{\{ i_0 + k^c[0] \geq d \}} \ctrl{3} & \gate[style={rounded corners, fill=gray!30}]{\{ i_0 + k^c[0] < d \}} \ctrl{3} & \gate[style={rounded corners, fill=gray!30}]{\{ i_0 + k^c[0] < d \}} \ctrl{3} & \qw & \qw \\
            & \lstick{$\ket{i_1}$} & \gate[style={rounded corners, fill=gray!30}]{\{ i_1 + k^c[1] \geq d-1 \}} \ctrl{2} & \gate[style={rounded corners, fill=gray!30}]{\{ i_1 + k^c[1] < d-1 \}} \ctrl{2} & \gate[style={rounded corners, fill=gray!30}]{\{ i_1 + k^c[1] \geq d \}} \ctrl{2} & \gate[style={rounded corners, fill=gray!30}]{\{ i_1 + k^c[1] < d \}} \ctrl{2} & \qw & \qw \\
            & \lstick{$\ket{i_2}$} & \gate[style={rounded corners, fill=gray!30}]{\{ i_2 + k^c[2] \geq d-1 \}} \ctrl{1} & \gate[style={rounded corners, fill=gray!30}]{\{ i_2 + k^c[2] \geq d \}} \ctrl{1} & \gate[style={rounded corners, fill=gray!30}]{\{ i_2 + k^c[2] \geq d-1 \}} \ctrl{1} & \gate[style={rounded corners, fill=gray!30}]{\{ i_2 + k^c[2] \geq d \}} \ctrl{1} & \qw & \qw \\
            & \lstick{$\ket{a}$} & \gate{X} & \gate{X} & \gate{X} & \gate{X} & \ctrl{1} & \qw \\
            & \lstick{$\ket{c}$} & \qw & \qw & \qw & \qw & \gate{X} & \qw
    \end{quantikz}
    }%
    \caption{Three-qudit quantum comparator circuit}
    \label{subcirc:n=3}
    \end{subfigure}
\caption{Circuits for $\hat{C}_k$ comparator, given $\mathcal{O}(1)$ available ancilla carry qubits, for an (a) $n=2$ and (b) $n=3$ qudit register. The diagrams follow the same notation as in Fig. \ref{circ:type1comparator}. Note that while these choices for $\hat{C}_k$ only require one ancilla carry qubit $\ket{a}$ (which, as always, is initialised to $\ket{0}$), the resulting circuit is much deeper, with many gates containing multiple qudits acting as controls. Each of these gates encodes the logic for a particular scenario in which sums of the individual digits in $i$ and $k^c$ result in an overall output carry bit. For example, for second gate in (b) above, the classical sum $i_0 + k^c[0]$ is large enough to result in an output carry bit, but $i_1 + k^c[1]$, plus the input carry bit from index $0$, is not big large enough to give one. However, $i_2 + k^c[2]$ is, even without the input carry bit (if any `control set' in any gate is empty, the gate falls away). As before, the bit value of the most significant qubit is encoded in the comparator qubit $\ket{c}$ with the final controlled-$X$ gate, and all of the other gates are uncomputed after this controlled-$X$ gate (not shown above).}
\label{circ:type2comparator}
\end{figure}

A second approach to constructing $\hat{C}_k$ instead only requires $\mathcal{O}(1)$ ancilla carry qubits. Such a circuit subroutine may be preferable for quantum hardware with limited resources yet long coherence times. Figs. \ref{subcirc:n=2} and \ref{subcirc:n=3} outline the $n=2$ and $n=3$ cases, respectively; the subroutine for general $n$ can be found recursively. However, having fewer ancilla qubits comes at the expense of requiring a much deeper circuit: at least $\mathcal{O}(n^2)$, counted with respect to the controlled-$X$ gates with an arbitrary number of control qudits\footnote{Since it's unclear how to efficiently decompose multi-controlled qudit operations - as in Fig. \ref{circ:type2comparator} - into single- or double-control gates - as in Fig. \ref{circ:type1comparator} - a fair comparison between these two approaches' scaling isn't practicable yet.}. Recall that it is indeed possible to employ qudits, rather than qubits, as the `carry' ancillas.

\subsection{Payoff encoding subroutine}
\label{subsec:payoffencoding}

To construct $\hat{L}_f$, first recall that the base-$d$ digit representation of $i$, is

\begin{equation}
i = i_0 + d^1i_1 + \cdots + d^{n-2}i_{n-2} + d^{n-1}i_{n-1},
\label{subseceqn:i}
\end{equation}

\noindent
so, using Eqs. \ref{eqn:affinemap} and \ref{eqn:ftilde}, $c\tilde{f}(i) + s$ can be written as

\begin{equation*}
c\tilde{f}(i) + s = \frac{2c}{s_{d^n-1}-K} \times \max\left(0, \omega i_0 + \omega d^1i_1 + \cdots + \omega d^{n-1}i_{n-1} + S_{\text{min}} + \frac{1}{2}\omega - K\right) - c + \frac{\pi}{4}.
\end{equation*}

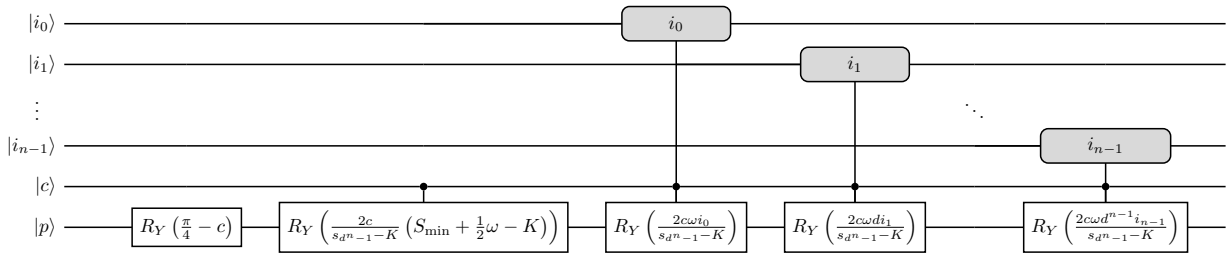
\begin{figure}[h]
    \centering
    \resizebox{\linewidth}{!}{%
    \begin{quantikz}[row sep={0.75cm,between origins}, column sep=0.7cm]
        & \lstick{$\ket{i_0}$} & \qw & \qw & \gate[style={rounded corners, fill=gray!30}]{i_0} \ctrl{5} & \qw & \qw & \qw & \qw \\
        & \lstick{$\ket{i_1}$} & \qw & \qw & \qw & \gate[style={rounded corners, fill=gray!30}]{i_1} \ctrl{4} & \qw & \qw & \qw \\
        & \vdots \hspace{1cm} &&&&& \ddots & & \\
        & \lstick{$\ket{i_{n-1}}$} & \qw & \qw & \qw & \qw & \qw & \gate[style={rounded corners, fill=gray!30}]{i_{n-1}} \ctrl{2} & \qw \\
        & \lstick{$\ket{c}$} & \qw & \ctrl{1} & \ctrl{1} & \ctrl{1} & \qw & \ctrl{1} & \qw \\
        & \lstick{$\ket{p}$} & \gate{R_Y \left( \frac{\pi}{4} - c \right)} & \gate{R_Y \left( \frac{2c}{s_{d^n-1}-K}\left( S_{\text{min}} + \frac{1}{2}\omega - K \right) \right)} & \gate{R_Y \left( \frac{2c \omega i_0}{s_{d^n-1}-K} \right)} & \gate{R_Y \left( \frac{2c \omega d i_1}{s_{d^n-1}-K} \right)} & \qw & \gate{R_Y \left( \frac{2c \omega d^{n-1} i_{n-1}}{s_{d^n-1}-K} \right)} & \qw
    \end{quantikz}
    }%
    \caption{$\hat{L}_f$ subroutine. Circuit for $\hat{L}_f$, with $\ket{p}$ the payoff qubit (initialised to $\ket{0}$) from step $2.2$ of section \ref{subsec:payoff}. The circuit follows a similar notation as in Figs. \ref{circ:type1comparator} and \ref{circ:type2comparator} (with any comparator subroutine ancillas $\ket{a}$ not shown here). However, here the blank grey `control bubbles' for each gate indicate which qudit is acting as the control, with the relationship between the control qudit's state and the $R_Y$ rotation angle given by the argument of the target rotation operation itself. For example, for qudit $0$, if $\ket{i_0} = \ket{0}$, then $R_Y \left( 0 \right) = I$ is applied to $\ket{p}$; if $\ket{i_0} = \ket{1}$, then $R_Y \left( \frac{2c \omega}{s_{d^n-1}-K} \right)$; if $\ket{i_0} = \ket{2}$, then $R_Y \left( \frac{4c \omega}{s_{d^n-1}-K} \right)$; etc., and so on for all qudits. However, these gates (except the very first one) are only applied if the state of the comparator qubit is $\ket{1}$ (this corresponds to the `max' function in the payoff $f$). Finally, note the different power of $d$ coefficients for each unique controlled rotation, which corresponds with the different digits in $i$, Eq. \ref{subseceqn:i}.}
    \label{circ:payoffload}
\end{figure}

The choice between the max function arguments is controlled by the comparator qubit. This relatively straightforward circuit is set out in Fig. \ref{circ:payoffload}. Since the mapping from $(\mathbb{R}, +)$ (the group of all real numbers, equipped with ordinary addition) to $(\text{R}_{\text{Y}}, \cdot)$ (the group of all $R_Y$ rotations, equipped with matrix multiplication) is a group homomorphism \cite{pinter2010book}, the contributions from each of the individual $Y$ rotations combine such that $R_Y(c\tilde{f}(i) + s)$ is ultimately the rotation applied to the payoff qubit.

\end{document}